\documentclass{article}
\usepackage{amsfonts}
\usepackage{amsmath}
\usepackage{graphicx}
\usepackage{cite}
\usepackage{hyperref}

\begin{document}

\title{Duffing-type equations: singular points of amplitude profiles and
bifurcations}
\author{Jan Kyzio\l , Andrzej Okni\'nski \\
Politechnika \'Swi\c{e}tokrzyska, Al. 1000-lecia PP 7, \\
25-314 Kielce, Poland}
\maketitle

\begin{abstract}
We study the Duffing equation and its generalizations with polynomial
nonlinearities. Recently, we have demonstrated that metamorphoses of the
amplitude response curves, computed by asymptotic methods in implicit form
as $F\left( \Omega ,\ A\right) =0$, permit prediction of qualitative changes
of dynamics occurring at singular points of the implicit curve $F\left(
\Omega ,\ A\right) =0$. In the present work we determine a global structure
of singular points of the amplitude profiles computing bifurcation sets,
i.e. sets containing all points in the parameter space for which the
amplitude profile has a singular point. We connect our work with independent research on 
tangential points on amplitude profiles, associated with jump phenomena, characteristic for 
the Duffing equation.  We also show that our techniques can
be applied to solutions of form $\Omega _{\pm }=f_{\pm }\left( A\right) $,
obtained within other asymptotic approaches.
\end{abstract}

\section{Introduction}
\label{Intro}

Recently, nonlinear Duffing-type oscillators attracted considerable attention
due to a rich variety of engineering applications. A very interesting class of
generalized Duffing oscillators with polynomial nonlinearities is described by the following equation:%
\begin{equation}
m\frac{d^{2}x}{dt^{2}}+\mu\frac{dx}{dt}+k_{1}x+k_{3}x^{3}+k_{5}x^{5}%
+\ldots+k_{n}x^{n}=a\cos\left(  \omega t\right)  ,\label{gen-Duff-1}%
\end{equation}
which, for example, can model dynamics of a mathematical pendulum or a ship-roll motions 
\cite{Kovacic2011,Kalmar2011,Mallik2011,Younesian2010,Karahan2017,Khatami2020,Hieu2018,Azimi2016,Zulli2016,Bikdash1994}.
In (\ref{gen-Duff-1}) $m$ is the mass, $\mu$ is the damping parameter, $k_{1}$
can be interpreted as the linear stiffness coefficient, $k_{n}\ \left(
n=3,5,\ldots\right)  $ are parameters of the nonlinear restoring force, while
$a$, $\omega$ are the amplitude and angular frequency of the periodic driving
force, respectively.

Nonlinear reponses of form $x\left( t\right) =A\left( \omega \right) \cos
\left( \omega t+\varphi \left( \omega \right) \right)$  can be computed by any of many 
asymptotic methods \cite{Nayfeh2011}. In the present work we compute the
asymptotic solutions  in implicit form as $F\left( \omega ,\ A,\text{\ }\underline{c}\right) =0$, where $\underline{c}
=\left( c_{1},c_{2},\ldots ,c_{m}\right) $ are parameters, 
by the application of the Krylov-Bogoliubov-Mitropolsky (KBM) method
\cite{Nayfeh2011}. 

As explained in Section \ref{General} and demonstrated in our earlier papers
qualitative changes of dynamics occur at singular points of the implicit
amplitude equation $F\left( \omega ,\ A,\text{\ }\underline{c}\right) =0$
(also known as the \ amplitude-frequency response  or the amplitude response
equation), see \cite{Kyziol2020} and references therein. Singular points
appear at some special values of parameters $\underline{c}=\underline{c}%
_{\ast }$, for which the implicit function $F$ fulfills appropriate
equations, see Eqs. (\ref{SINGULAR}). In other words, at $\underline{c}=%
\underline{c}_{\ast }$ there is a change of differential properties of the
amplitude response curve $F\left( \omega ,\ A,\text{\ }\underline{c}_{\ast
}\right) =0$ at a singular point $\left( \omega _{\ast },\ A_{\ast }\right) $, 
referred to as a metamorphosis.

In this work, we attempt to find a global picture
of singular points of the amplitude response equation for the generalized Duffing
equation (\ref{gen-Duff-1}). We derive formulae for singular points and the
bifurcation set in general case and apply these results to the case of the
standard cubic Duffing equation ($n=3$) and to the cubic-quintic Duffing
equation ($n=5$). 
We shall carry out this program for higher values of $n$ in our future work.

In Section \ref{Classification} we connect our work on singular metamorphoses with 
research on changes of differential properties of amplitude-frequency response curves 
in a non-singular case, which can be viewed as non-singular metamorphoses 
\cite{Kalmar2011}. Metamorphoses described in \cite{Kalmar2011} can be 
classified as vertical tangencies -- for some parameter values the amplitude response 
curve has vertical tangent points (critical but non-singular)  associated with saddle-node bifurcations 
(jump phenomena) \cite{Kalmar2011}.

In \ref{A} we analyse metamorphoses of the amplitude profiles from another point of view. 
Implicit equation  $F\left( \Omega ,\ A\right) =0$ can be solved for $\Omega$ for the generalized 
Duffing equation so that we can get explicit solutions $\Omega _{\pm }=f_{\pm }\left( A\right)$.
 Intersection conditions for these two branches, $f_{+}\left( A\right) =f_{-}\left( A\right)$, 
yield singular points of the amplitude profile 
(the same as described in Section \ref{examples}, 
obtained within the more general approach of Section \ref{B-set}). 
We present a more detailed picture of transformation of a non-singular amplitude profile 
into a self-intersection, non-singular amplitude, and into amplitude with an isolated point.

We also apply our approach to explicit solutions  $\omega _{\pm
}=g_{\pm }\left( A\right) $ computed in \cite{Karahan2017} for the
cubic-quintic oscillator obtained via the Multiple Scales Linstedt Poincar\'{e} (MSLP) method, 
see \ref{B}. We show that our techniques, described in
Section \ref{Global} as well as in the \ref{A}), can be applied to amplitudes obtained
within another asymptotic approach.

Finally, in \ref{Details} computational details are described.

\section{The Krylov-Bogoliubov-Mitropolsky amplitude profiles for the generalized Duffing equation}
\label{KBM}

Introducing nondimensional units, $\Omega=\dfrac{\omega}{\omega_{n}}$,
$\omega_{n}=\sqrt{\dfrac{k_{1}}{m}}$, $\tau=\omega_{n}t$, Eq.
(\ref{gen-Duff-1}) is cast into form:
\begin{equation}
\ddot{y}+h\dot{y}+y+c_{3}y^{3}+c_{5}y^{5}+\ldots+c_{n}y^{n}=f\cos\left(
\Omega\tau\right)  , \label{gen-Duff-2}%
\end{equation}
where overdots denote derivatives with respect to $\tau$ and $n=3,5,7,9,\ldots$ .
We assume that parameters $h$, $c_{3}$, \ldots , $c_{n}$, $f$ are small and
can be thus written as $h=\varepsilon \bar{h}$, $c_{3}=\varepsilon \bar{c}%
_{3}$, \ldots , $c_{n}=\varepsilon \bar{c}_{n}$, $f=\varepsilon \bar{f}$
where $\varepsilon $ is a small parameter and $\bar{h}$, $\bar{c}_{3}$, \ldots
, $\bar{c}_{n}$, $\bar{f}$ are of order $O\left( 1\right) $. 

Looking for $1:1$ resonances we rewrite equation \ref{gen-Duff-2} in form of
a weakly perturbed system:%
\begin{equation}
\frac{d^{2}y}{d\tau ^{2}}+\Omega ^{2}y+\varepsilon \left( \sigma y+g\right)
=0,  \label{prepared}
\end{equation}%
where 
\begin{subequations}
\label{PERT}
\begin{eqnarray}
g &=&\bar{h}\frac{dy}{d\tau }+\bar{c}_{3}y^{3}+\bar{c}_{5}y^{5}+\ldots +\bar{%
c}_{n}y^{n}-\bar{f}\cos \left( \Omega \tau \right) ,  \label{perturbation} \\
\varepsilon \sigma  &=&1-\Omega ^{2},  \label{detuning}
\end{eqnarray}
\end{subequations}
and $\sigma =O\left( 1\right) $.

We apply the Krylov-Bogoliubov-Mitropolsky (KBM) perturbation approach \cite{Nayfeh2011} assuming 
the  $1:1$ resonance  in form:
\begin{equation}
y=A\left( \tau \right) \cos \left( \Omega \tau +\varphi \left( \tau \right)
\right) +\varepsilon y_{1}\left( A,\varphi ,\tau \right) +\ldots
\label{soly}
\end{equation}
with slowly varying amplitude $A$ and phase $\varphi $: 
\begin{subequations}
\label{SECULAR1}
\begin{eqnarray}
\frac{dA}{d\tau } &=&\varepsilon M_{1}\left( A,\varphi \right) +\ldots ,
\label{Am} \\
\frac{d\varphi }{d\tau } &=&\varepsilon N_{1}\left( A,\varphi \right)
+\ldots ,  \label{phi}
\end{eqnarray}
\end{subequations}
obtaining:
\begin{subequations}
\label{SECULAR2}
\begin{eqnarray}
M_{1} &=&\frac{1}{2\Omega }\left( -\bar{h}A\Omega -\bar{f}\sin \varphi
\right) ,  \label{sec1} \\
N_{1} &=&\frac{1}{2\Omega A}\left( \sigma A+\bar{d}_{3}A^{3}+\bar{d}%
_{5}A^{5}+\ldots \bar{d}_{n}-\bar{f}\cos \varphi \right) ,  \label{sec2}
\end{eqnarray}
\end{subequations}
with $\bar{d}_{n}$  given in (\ref{d}).

The fixed points of the slow-flow equations (\ref{SECULAR1}),  (\ref{SECULAR2}) correspond to solutions with
constant amplitude and phase \cite{Kalmar2011}.
We thus demand that $M_1 = N_1 = 0$, eliminate $\varphi$, and get the following implicit 
amplitude-frequency equation:
\begin{equation}
\begin{array}{l}
L_{n}\left( X,Y\right) =h^{2}XY+Y\left( X-1-d_{3}Y-d_{5}Y^{2}-\ldots
-d_{n}Y^{n}\right) ^{2}-f^{2}=0\smallskip  \\ 
X\equiv \Omega ^{2},\ Y\equiv A^{2}%
\end{array}
\label{Amplitude}
\end{equation}
and 
\begin{equation}
\bar{d}_{n}=2^{\frac{1-n}{2}}\frac{n!!}{\left( \frac{n+1}{2}\right) !}\ \bar{%
c}_{n},\quad n=3,5,7,9,\ldots .   \label{d}
\end{equation}
In what follows we shall also write $F\left( \Omega ,A\right) =L_n\left(
\Omega ^{2},A^{2}\right) $.

\section{Singular points of amplitude profiles and bifurcations of
dynamics}
\label{General}
Detailed description of properties and applications of singular points of
amplitude profiles can be found in Ref. \cite{Kyziol2020}. In short, solving
a nonlinear differential equation of form: 
\begin{equation}
\frac{d^{2}y}{d\tau ^{2}}+\omega ^{2}y=\varepsilon f\left( y,\frac{dy}{d\tau 
},\tau \right) ,  \label{ODE}
\end{equation}%
where $\varepsilon $ is a small parameter and $f$ is a periodic function of
time $\tau $ with period $T=\frac{2\pi }{\Omega }$, by an asymptotic method
we find an approximate solution:
\begin{equation}
y\left( \tau \right) =A\left( \Omega \right) \cos \left( \Omega \tau
+\varphi \left( \Omega \right) \right) +\varepsilon y_{1}\left( \tau \right)
+\ldots   \label{Solution}
\end{equation}
where the amplitude $A$ and frequency $\Omega$ fulfill the
amplitude response equation:
\begin{equation}
F\left(  \Omega, A;\underline{c}\right)  =0, \label{AE}
\end{equation}
where $\underline{c}=\left(  c_{1},c_{2},\ldots,c_{m}\right)  $ are
parameters. Equation (\ref{AE}) defines an implicit function -- a
two-dimensional planar curve - the amplitude-frequency response curve (the amplitude profile). 
The form of this curve,
as well as stability of the solution (\ref{Solution}), determine
(approximately) dynamics of the system.

Qualitative changes of shape of the amplitude profile
(\ref{AE}), which are equivalent to changes of differential properties of these 
curves and are 
 referred henceforth as metamorphoses, induced by smooth changes of
control parameters $\underline{c}$, lead to qualitative changes of dynamics
(bifurcations). 
According to the differential geometry of curves \cite{Spivak1965,Wall2004}%
\ an implicit curve changes its form at singular points which fulfill the
following equations:
\begin{subequations}
\label{SINGULAR}
\begin{align}
F\left( \Omega ,A;\underline{c}\right) & =0,  \label{S1} \\
\frac{\partial F\left( \Omega ,A;\underline{c}\right) }{\partial \Omega }&
=0,  \label{S2} \\
\frac{\partial F\left( \Omega ,A;\underline{c}\right) }{\partial A}& =0.
\label{S3}
\end{align}
\end{subequations}
Solutions of Eq. (\ref{SINGULAR}), if exist, are of form $\Omega =\Omega
_{\ast }$, $A=A_{\ast }$, $\underline{c}=\underline{c}_{\ast }$.
Accordingly, the amplitude response curve $F\left( \Omega ,A,\underline{c}%
_{\ast }\right) =0$ changes its differential properties at singular point $%
\left( \Omega _{\ast },A_{\ast }\right) $. 

Metamorphoses of the amplitude-frequency curves (i.e. changes of differential properties) 
also can occur in a non-singular setting. More precisely, a metamorphosis of this kind occurs when 
a smooth change of parameters  $\underline{c}$ leads to formation of vertical tangent points 
of an amplitude profile. This gives rise to the so-called jump phenomenon, first described 
in the context of change of differential properties of the amplitude response curve for 
the cubic Duffing equation in \cite{Kalmar2011}. It follows that equations guaranteeing 
formation of a (non-singular) vertical tangent point  $\left( \Omega _{\ast },A_{\ast }\right) $  are:
\begin{subequations}
\label{NONSINGULAR}
\begin{align}
F\left( \Omega ,A;\underline{c}\right) & =0,  \label{NS1} \\
\frac{\partial F\left( \Omega ,A;\underline{c}\right) }{\partial A}& =0,
\label{NS3}
\end{align}
\end{subequations}
see Section \ref{Classification} for more details.

Investigation of metamorphoses  of 
amplitude profiles induced by change of parameters was carried out 
in the framework of Catastrophe Theory 
in \cite{Holmes1976}\ for the Duffing equation in a non-singular case.
The idea to use Implicit Function Theorem to
\textquotedblright define and find different branches intersecting at singular
points\textquotedblright\ of amplitude profiles was proposed in
\cite{Awrejcewicz1995}.

While changes of differential properties of asymptotic solutions are important, stability of 
the solutions is another essential factor shaping the dynamics. Stability of the slow-flow 
equations (\ref{SECULAR1}) is determined by eigenvalues of the Jacobian matrix \cite{Nayfeh2008}:
\begin{equation}
\mathbb{J}=\left( 
\begin{array}{cc}
\dfrac{\partial M_{1}}{\partial A} & \dfrac{\partial M_{1}}{\partial \varphi 
}\medskip  \\ 
\dfrac{\partial N_{1}}{\partial A} & \dfrac{\partial N_{1}}{\partial \varphi 
}%
\end{array}%
\right)   \label{J}
\end{equation}
We show in Section \ref{Classification} that changes of differential properties of 
asymptotic solutions and changes of their stability (bifurcations) are strictly related.
\section{Global view of metamorphoses of the amplitude profiles: general
case and examples for $n=3,\ 5$}
\label{Global}
\subsection{Singular points}
We shall investigate singular points of the amplitude equation (\ref%
{Amplitude}) because bifurcations occur at these points, cf. Section \ref
{General}. Singular points of algebraic curve $L_{n}\left( X,Y\right) =0$
are given by equations: 
\begin{subequations}
\label{SING1}
\begin{align}
L_{n}\left( X,Y\right) & =0,  \label{Sing1a} \\
\tfrac{\partial L_{n}\left( X,Y\right) }{\partial X}& =0,  \label{Sing1b} \\
\tfrac{\partial L_{n}\left( X,Y\right) }{\partial Y}& =0.  \label{Sing1c}
\end{align}
\end{subequations}
Eqs. (\ref{SING1}) can be solved for a general odd integer $n\geq 3$: 
\begin{equation}
\left. 
\begin{array}{l}
X=\tfrac{1}{2}-\tfrac{3}{8}h^{2}-\tfrac{1}{2}d_{5}Z^{2}-d_{7}Z^{3}-\tfrac{3}{%
2}d_{9}Z^{4}-2d_{11}Z^{5}-\ldots -\tfrac{n-3}{4}d_{n}Z^{\frac{n-1}{2}} \\ 
Y=Z \\ 
d_{3}=\tfrac{-4+h^{2}}{8Z}-\left( \tfrac{3}{2}d_{5}Z+2d_{7}Z^{2}+\tfrac{5}{2}%
d_{9}Z^{3}+3d_{11}Z^{4}+\ldots +\left( 1+\tfrac{n-3}{4}\right) d_{n}Z^{\frac{%
n-3}{2}}\right) 
\end{array}%
\right\} ,  \label{solution}
\end{equation}%
where $Z$ is a solution of the polynomial equation $g_{n}\left( Z\right) =0$:%
\begin{equation}
\left. 
\begin{array}{l}
g_{n}\left( Z\right) =2\left( n-1\right) h^{2}d_{n}Z^{\frac{n+1}{2}}+\ldots
+16h^{2}d_{11}Z^{6}+12h^{2}d_{9}Z^{5}+8h^{2}d_{7}Z^{4} \\ 
+4h^{2}d_{5}Z^{3}+\left( h^{4}-4h^{2}\right) Z+8f^{2}=0%
\end{array}%
\right.   \label{gn(Z)}
\end{equation}

\subsection{Bifurcation set}
\label{B-set}

It follows from general theory of implicit functions that in a singular
point there are multiple solutions of equation (\ref{Amplitude}) \cite{Kyziol2019,Kyziol2020}. We shall use
this property to compute parameters values for which the amplitude profile
defined by equation (\ref{Amplitude}) has singular points. We shall refer to such
set in the parameter space as the bifurcation set, see Ref. \cite{Holmes1976}
where this term was used in the context of multiple solutions of the
amplitude equation for the Duffing equation in the non-singular case. 

To define a singular point we can use equation (\ref{Sing1a}), and Eq. (\ref%
{Sing1b}) which excludes existence of the single-valued function $X=f\left(
Y\right) $ and an alternative to condition (\ref{Sing1c}) which excludes
existence of the single-valued function $Y=g\left( X\right) $. 

We thus solve  Eqs. (\ref{Sing1a}), (\ref{Sing1b}) 
\begin{subequations}
\label{BIF1}
\begin{align}
L_{n}\left( X,Y\right) & =0,  \label{b1} \\
\tfrac{\partial L_{n}\left( X,Y\right) }{\partial X}& =0,  \label{b2}
\end{align}
\end{subequations}
obtaining: 
\begin{subequations}
\begin{align}
X& =1-\tfrac{1}{2}h^{2}+d_{3}Z+d_{5}Z^{2}+d_{7}Z^{3}+d_{9}Z^{4}+d_{11}Z^{5}+%
\ldots +d_{n}Z^{\frac{n-1}{2}}  \label{X} \\
Y& =Z  \label{Y}
\end{align}
where $Z$ is a solution of the polynomial equation $f_{n}\left( Z\right) =0$:
\begin{equation}
\left. 
\begin{array}{l}
f_{n}\left( Z\right) =h^{2}d_{n}Z^{\frac{n+1}{2}}+\ldots
+h^{2}d_{11}Z^{6}+h^{2}d_{9}Z^{5}+h^{2}d_{7}Z^{4}+h^{2}d_{5}Z^{3} \\ 
+h^{2}d_{3}Z^{2}+\left( 4h^{2}-h^{4}\right) Z-f^{2}=0.
\end{array}%
\right.   \label{fn(Z)}
\end{equation}
\end{subequations}
Note that roots of the polynomial $f_{n}$ are values of the amplitude
function $Y=g(X)$ in critical points (i.e. at maxima, minima or inflexion
points). Indeed, suppose that $Y=g\left( X\right) $ is a solution of Eq. (%
\ref{b1}) and $\frac{\partial L_{n}\left( X,Y\right) }{\partial Y}\neq 0$.
Then we have $\frac{dY}{dX}=g^{\prime }\left( X\right) =-\frac{\partial
L_{n}/\partial X}{\partial L_{n}/\partial Y}$ and it follows from Eq. (\ref%
{b2} that $g^{\prime }\left( X\right) =0$. We show that critical points
shape bifurcation diagrams.

We now demand that there are multiple solutions of equation (\ref{fn(Z)}) -- 
this conditions, an alternative to Eq. (\ref{Sing1c}), guarantees singularity of solutions 
of Eq. (\ref{Sing1a}).
Necessary and suficient condition for a polynomial to have multiple roots is
that its discriminant $\Delta$ vanishes \cite{Gelfand2008}, see also lecture
notes \cite{Janson2010}. Discriminant $\Delta$ can be computed as a
resultant of a polynomial $f\left( X\right) $ and its derivative $f^{\prime} 
$, with a suitable normalizing factor.

Polynomials $f$ and $g$ have a common root if and only if their resultant is
zero. More exactly, resultant $R\left( f,g\right) $ of two polynomials, $%
f\left( X\right) =a_{n}X^{n}+\ldots +a_{1}X+a_{0}$, $g\left( X\right)
=b_{m}X^{m}+\ldots +b_{1}X+b_{0}$, is given by determinant of the $\left(
m+n\right) \times \left( m+n\right) $ Sylvester matrix -- see, for example, Eq. (1) in \cite{Janson2010}:

\begin{equation}
R\left( f,g\right) =\det \left( 
\begin{array}{ccccccc}
a_{n} & a_{n-1} & a_{n-2} & \ldots  & 0 & 0 & 0 \\ 
0 & a_{n} & a_{n-1} & \ldots  & 0 & 0 & 0 \\ 
\vdots  & \vdots  & \vdots  &  & \vdots  & \vdots  & \vdots  \\ 
0 & 0 & 0 & \ldots  & a_{1} & a_{0} & 0 \\ 
0 & 0 & 0 & \ldots  & a_{2} & a_{1} & a_{0} \\ 
b_{m} & b_{m-1} & b_{m-2} & \ldots  & 0 & 0 & 0 \\ 
0 & b_{m} & b_{m-1} & \ldots  & 0 & 0 & 0 \\ 
\vdots  & \vdots  & \vdots  &  & \vdots  & \vdots  & \vdots  \\ 
0 & 0 & 0 & \ldots  & b_{1} & b_{0} & 0 \\ 
0 & 0 & 0 & \ldots  & b_{2} & b_{1} & b_{0}%
\end{array}%
\right) .  \label{resultant}
\end{equation}%

Therefore, the bifurcation set $\mathcal{M}\left( f,h,c_{3},c_{5},\ldots
,c_{n}\right) $ for a generalized Duffing equation (\ref{gen-Duff-2}) reads:%
\begin{equation}
R\left( f_{n},f_{n}^{\prime }\right) =0,  \label{bifurcation set}
\end{equation}
where the polynomial $f_{n}\left( Z\right) $ is defined in Eq. (\ref{fn(Z)}%
), and parameters $d_{n}$ are given in Eq. (\ref{d}). 

\subsection{Bifurcation sets for the cubic and cubic-quintic Duffing
equations}
\label{examples}

Let us first consider the cubic-qiuntic Duffing equation since results for
the cubic Duffing equation can be easily obtained by setting $c_{5}=0$. For
the cubic-quintic Duffing equation we get from Eq. (\ref{Amplitude}):
\begin{equation}
\begin{array}{l}
L_{5}\left( X,Y\right) =h^{2}XY+Y\left( X-1-d_{3}Y-d_{5}Y^{2}\right)
^{2}-f^{2}=0, \\ 
d_{3}=\tfrac{3}{4}c_{3},\ d_{5}=\tfrac{5}{8}c_{5},\quad \left( X\equiv
\Omega ^{2},\ Y\equiv A^{2}\right) 
\end{array}
\label{L(3-5)}
\end{equation}
and it follows from Eq. (\ref{fn(Z)}) that we have to consider conditions
for multiple roots for the polynomial:%
\begin{equation}
\begin{array}{l}
f_{5}\left( Z\right) =aZ^{3}+bZ^{2}+cZ+d=0, \\ 
a=5h^{2}c_{5},\ b=6h^{2}c_{3},\ c=\left( -2h^{4}+8h^{2}\right) ,\ d=-8f^{2}.%
\end{array}
\label{f(Z)}
\end{equation}

Condition for multiple roots is (see Eqs. (\ref{resultant}), (\ref%
{bifurcation set})):%
\begin{equation}
\begin{array}{l}
R\left( f_{5},f_{5}^{\prime }\right) =\det \left( 
\begin{array}{ccccc}
a & b & c & d & 0 \\ 
0 & a & b & c & d \\ 
3a & 2b & c & 0 & 0 \\ 
0 & 3a & 2b & c & 0 \\ 
0 & 0 & 3a & 2b & c%
\end{array}%
\right) = \\ 
=a\left( 4ac^{3}-b^{2}c^{2}+4db^{3}-18abdc+27a^{2}d^{2}\right) =0.%
\end{array}
\label{multiple1}
\end{equation}

Since $a\neq 0$ condition for multiple (double and triple) roots is:

\begin{equation}
4ac^{3}-b^{2}c^{2}+4db^{3}-18abdc+27a^{2}d^{2}=0,  \label{multiple2}
\end{equation}
or, after substituting expressions (\ref{f(Z)}) for $a,b,c,d$ we get
equation defining the bifurcation set $\mathcal{M}\left(
f,h,c_{3},c_{5}\right) $:

\begin{equation}
\begin{array}{l}
10c_{5}h^{10}+\left( 9c_{3}^{2}-120c_{5}\right) h^{8}+\left(
-72c_{3}^{2}+480c_{5}\right) h^{6} \\ 
+\left( 540c_{5}c_{3}f^{2}+144c_{3}^{2}-640c_{5}\right) h^{4} \\ 
+\left( -2160c_{5}+432c_{3}^{2}\right) c_{3}f^{2}h^{2}-2700c_{5}^{2}f^{4}=0.%
\end{array}
\label{bifset}
\end{equation}

\medskip It is now possible to find condition for degenerate singular
points. A cubic polynomial (\ref{f(Z)}) has a triple root if, apart from the
condition (\ref{multiple1}), also $f$ and $f^{\prime \prime }$ have a common
root:%
\begin{equation}
R\left( f_{5},f_{5}^{\prime \prime }\right) =\det \left( 
\begin{array}{cccc}
a & b & c & d \\ 
6a & 2b & 0 & 0 \\ 
0 & 6a & 2b & 0 \\ 
0 & 0 & 6a & 2b%
\end{array}%
\right) =-8a\left( 2b^{3}-9abc+27da^{2}\right) =0.  \label{triple1}
\end{equation}
Therefore, conditon for a triple root reads:
\begin{align}
4ac^{3}-b^{2}c^{2}+4db^{3}-18abdc+27a^{2}d^{2}& =0  \label{triple2} \\
2b^{3}-9abc+27da^{2}& =0  \notag
\end{align}%
and, after expressions for $a,b,c,d$ (\ref{f(Z)}) are invoked, the solution,
defining degenerate bifurcation set $\mathcal{M}_{\text{deg}}\left( f,h,c_{3},c_{5}\right)$, is:
\begin{equation}
f=\pm \dfrac{1}{6c_{3}}\sqrt{-c_{3}}h\left( h^{2}-4\right) ,\quad c_{5}=%
\dfrac{6c_{3}^{2}}{5\left( 4-h^{2}\right) }.  \label{degbifset}
\end{equation}
We can now easily extract expression for the bifurcation set for
the cubic Duffing equation. Indeed, substituting in Eq. (\ref{bifset}) $%
c_{5}=0$, we get:%
\begin{equation}
9c_{3}^{2}h^{2}\left( h^{6}-8h^{4}+16h^{2}+48c_{3}f^{2}\right) =0,
\label{bifset-cubic1}
\end{equation}
and thus the bifurcation set $\mathcal{M}\left( f,h,c_{3}\right) $ is:
\begin{equation}
c_{3}=-\frac{\left( h^{2}-4\right) ^{2}}{48f^{2}}h^{2}\quad \left( c_{3}\neq
0,\ h\neq 0,\ f\neq 0\right) .  \label{bifset-cubic2}
\end{equation}
Moreover, it follows from (\ref{degbifset}) that there are no degenerate
singular points for the cubic Duffing equation.
\section{Singular points of the amplitude profile for the cubic Duffing equation and 
the corresponding bifurcations}
\label{cubic} 
Substituting in Eq. (\ref{bifset-cubic2}) $h=1$, $f=1$ we compute $
c_{3}=-0.1875$ (case of the softening spring) and $X=\Omega ^{2}=\frac{1}{8}$, $Y=A^{2}=\frac{8}{3}$. 
\begin{figure}[h!]
\center
\includegraphics[width=6cm, height=6cm]{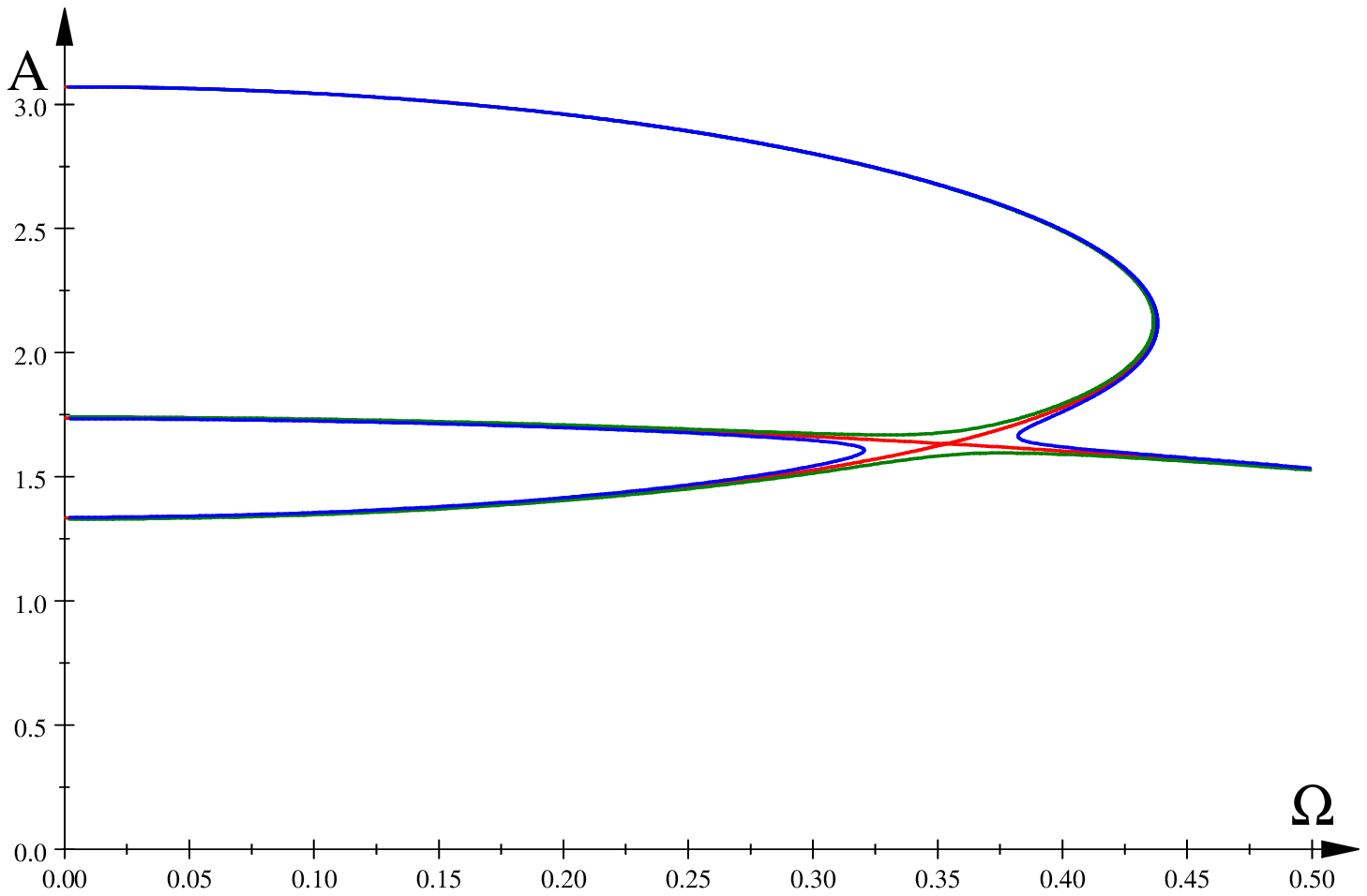}
\includegraphics[width=6cm, height=6cm]{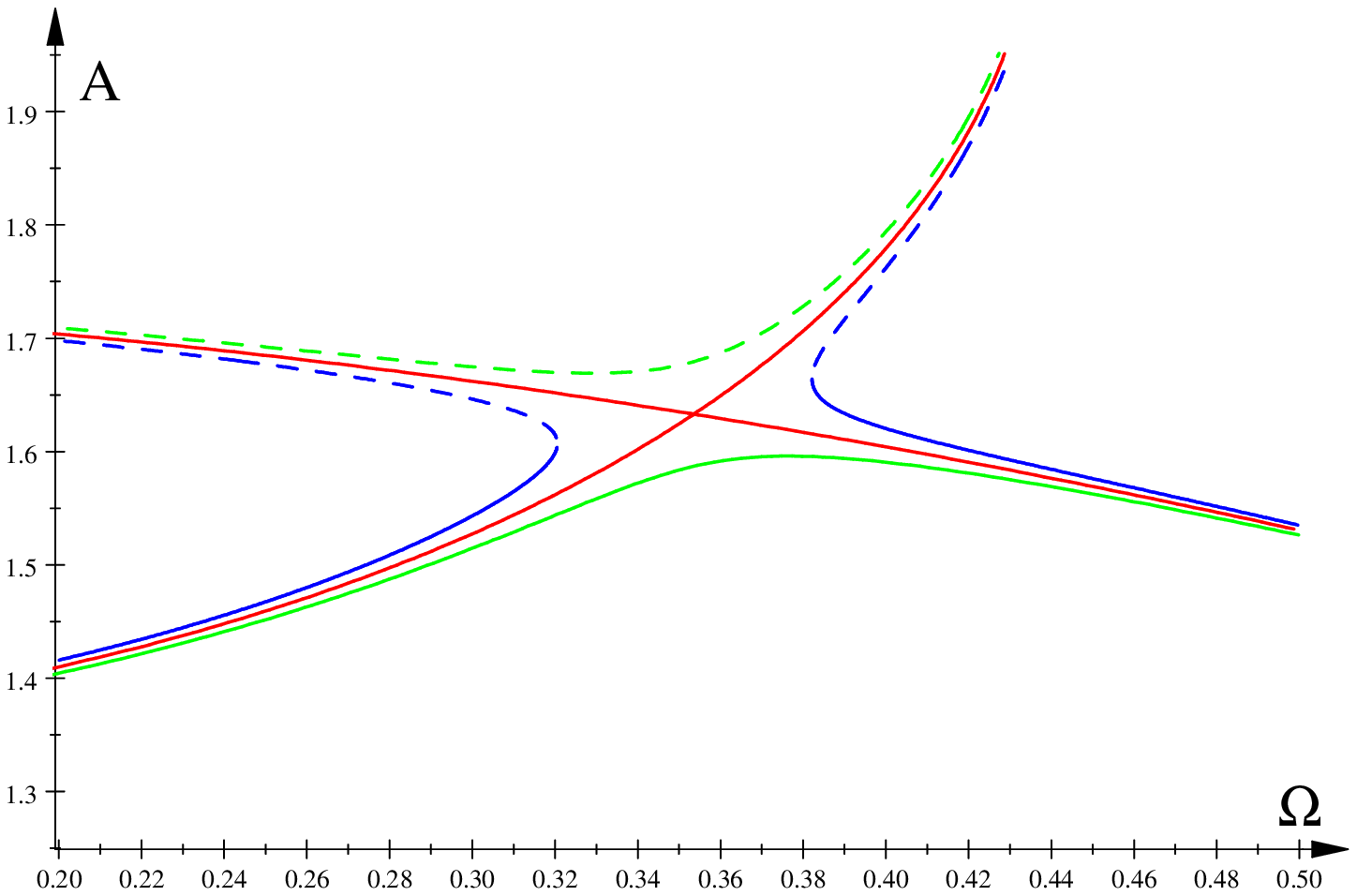}
\caption{Amplitude profiles, $f=1$ ( singular, red), $f=0.999$ (green), and $f=1.001$ (blue) -- left, neighborhood of the singular point, unstable branches marked by dashed  lines -- right.  }
\label{F1}
\end{figure}
\begin{figure}[h!]
\includegraphics[width=6cm, height=6cm]{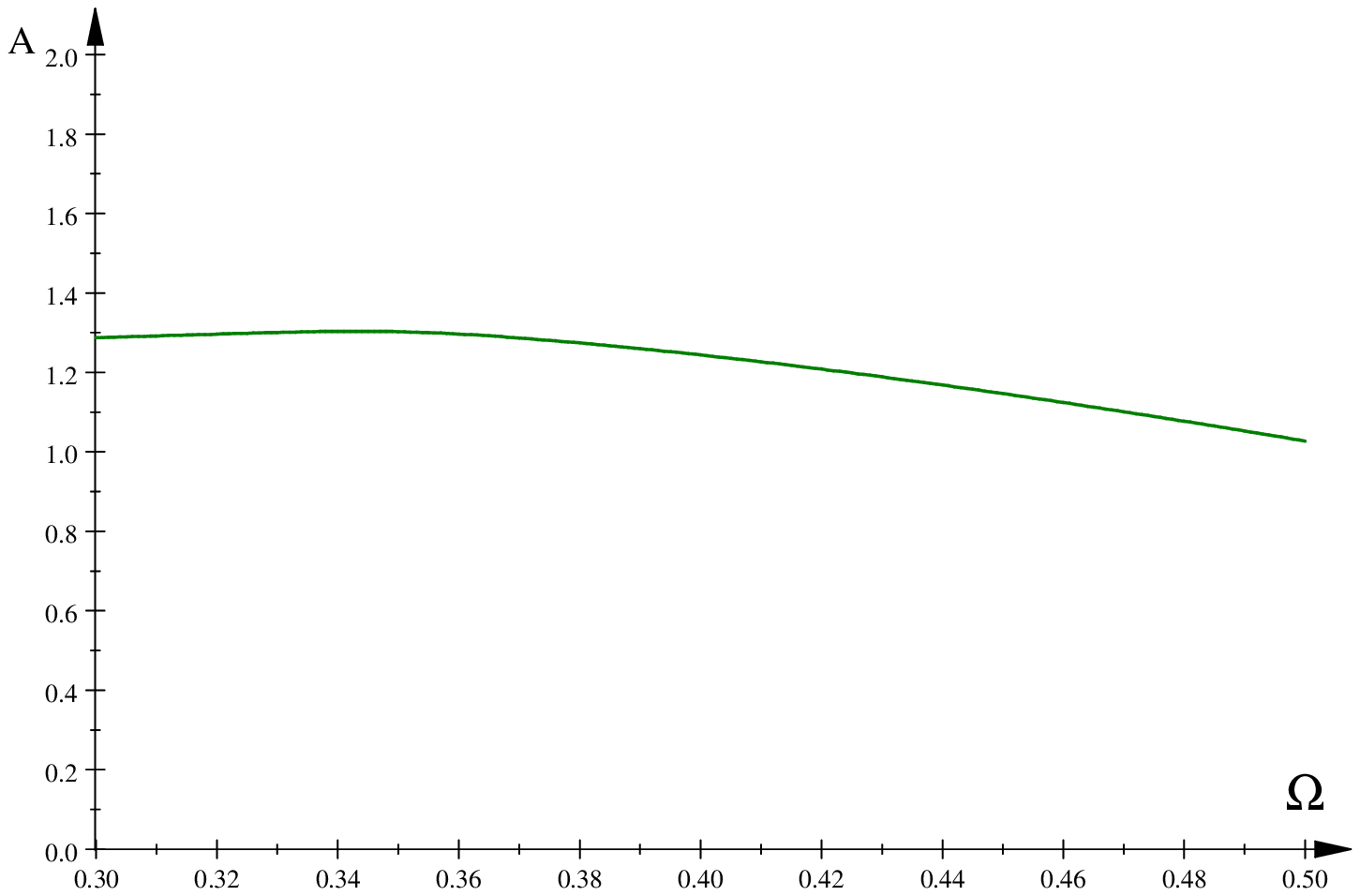}
\includegraphics[width=6cm, height=6cm]{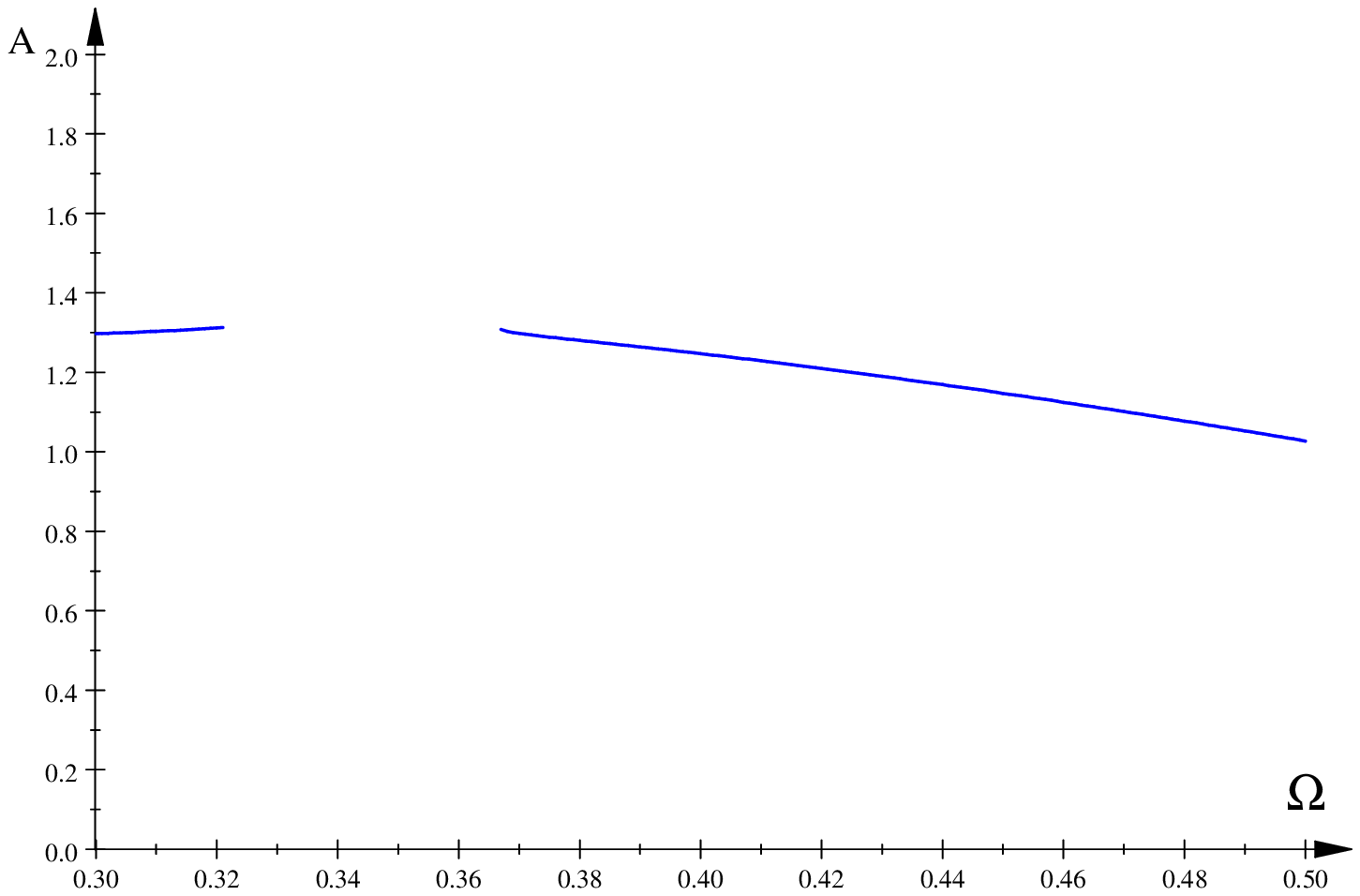}
\caption{Bifurcation diagrams, $f=1.060$, left (green) and $f=1.061$, right (blue)}
\label{F2}
\end{figure}

The
corresponding amplitude profile, as well as two non-singular curves, are
shown in Fig. \ref{F1}.  
The singular curve (red) corresponds to self-intersection. The corresponding 
bifurcation diagrams, computed by numerical integration of the differential equation (\ref{gen-Duff-2}), $n=3$,  
 are  shown in Fig. \ref{F2}.

 It follows that there is indeed a gap on the bifurcation diagram (blue), 
corresponding to discontinuity on the amplitude profile (blue).  In the case of numerical computation 
the discontinuity occurred for $1.060<f<1.061$,  in good agreement with the  predicted value $f = 1$.

\section{Examples of bifurcations at singular points of the amplitude profile
for the cubic-quintic Duffing equation}
\label{cubic-quintic}
It follows from Eqs. (\ref{degbifset}) that the cubic-quintic equation has a degenerate singular 
point for $c_3<0$. Moreover, in the neighbourhood of this point 
there are two families of non-degenerate singular points: isolated points and self-intersections. 

Choosing, for example, $h=0.2$, $c_{3}=-0.2$, we compute from ((\ref{degbifset}) 
other parameters of the degenerate singular point as $c_{5}=1.\,212\,121\times 10^{-2}$, $f=0.295\,161$ 
and $\Omega =0.565\,685$, $A=2.\,569\,047$. 

Now, for  $h=0.2$, $c_{3}=-0.2$ and $ c_{5}=1.15\times 10^{-2}$, 
we compute from Eqs. (\ref{SING1}) for $n=5$: $f=0.282\,240$ -- an isolated 
singular point with $\Omega =0.472\,685$, $A=2.\,920\,851$ 
and  $f=0.290\,089$ -- a self-intersection with $\Omega =0.617\,184$, $A=2.\,319\,843$. 
\begin{figure}[h!]
\center
\includegraphics[width=10.5cm, height=7cm]{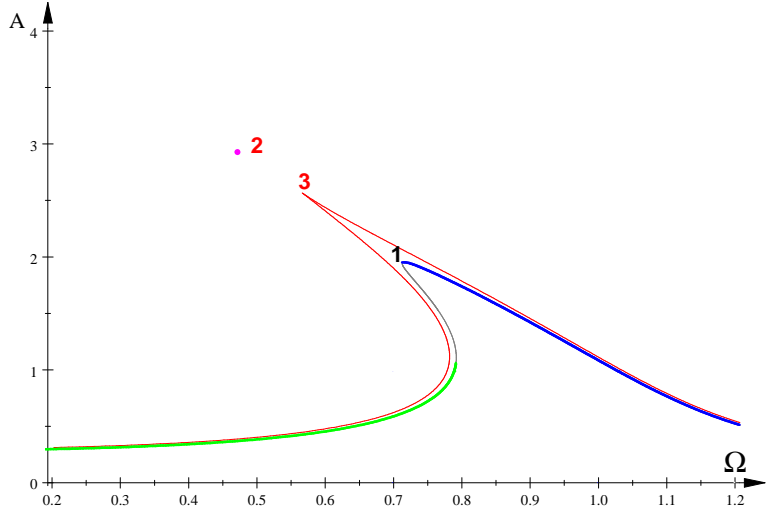}
\caption{Amplitude profiles: degenerate, $c_{5}=1.\,212\,121\times 10^{-2}, f=0.295\,161$ (red), 
isolated point, $c_{5}=1.15\times 10^{-2},$ $f=0.282\,240$ (green, blue, magenta)}
\label{F3}
\end{figure}
The corresponding amplitude profiles  for the degenerate and the isolated point 
are shown in Fig. \ref{F3}. Green, blue and magenta colours have been used to show 
correspondence with bifurcation diagrams computed numerically near the isolated point, see Fig. \ref{F4}. 
In the case of numerical integration of the Duffing equation (\ref{gen-Duff-2}), see \ref{Details} for 
computational details, an isolated point appears for $0.2990<f<0.2991$ in good agreement with the 
predicted value ($f=0.282\,240$).

\begin{figure}[t!]
\center
\includegraphics[width=6cm, height=6cm]{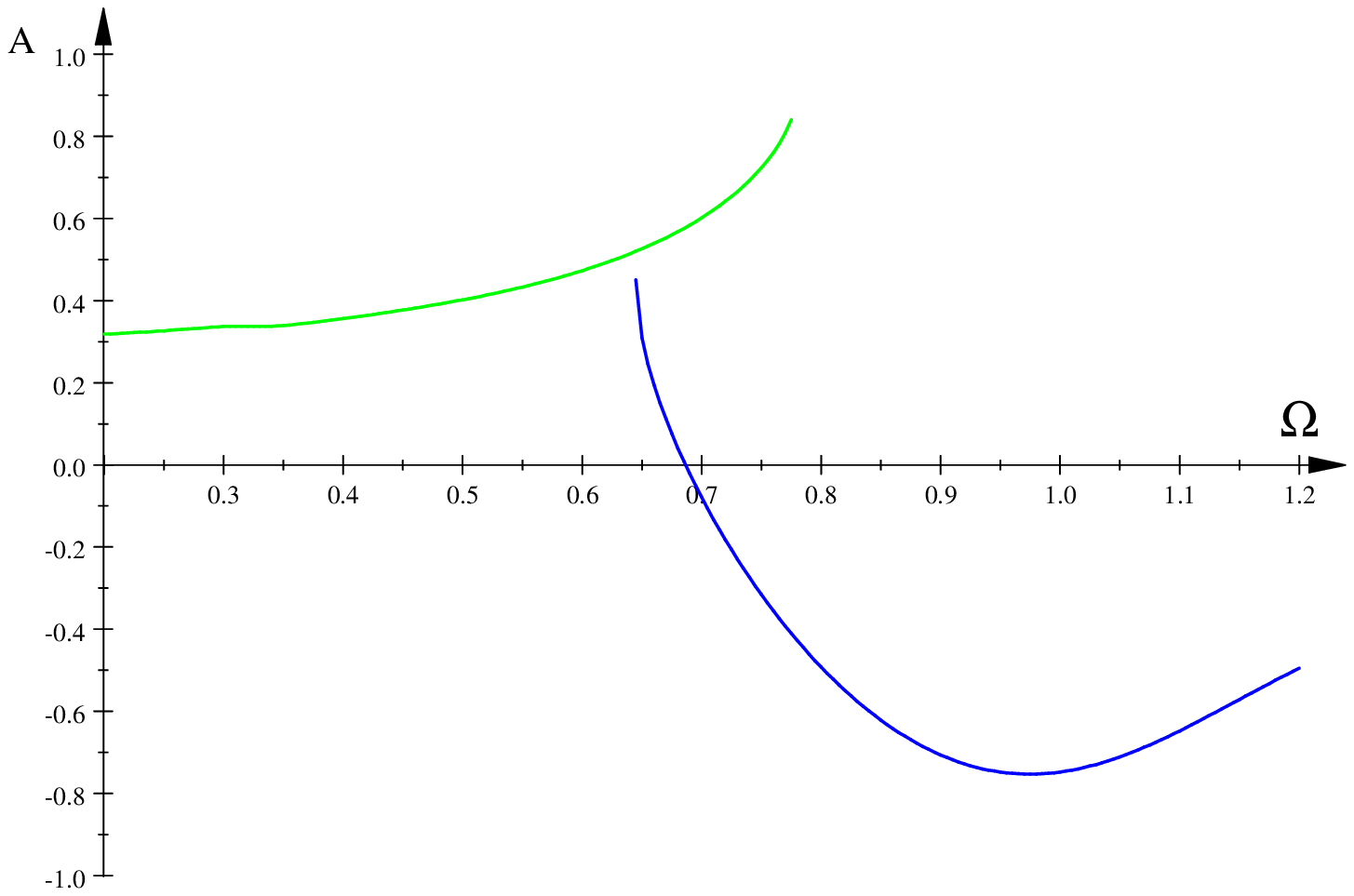}
\includegraphics[width=6cm, height=6cm]{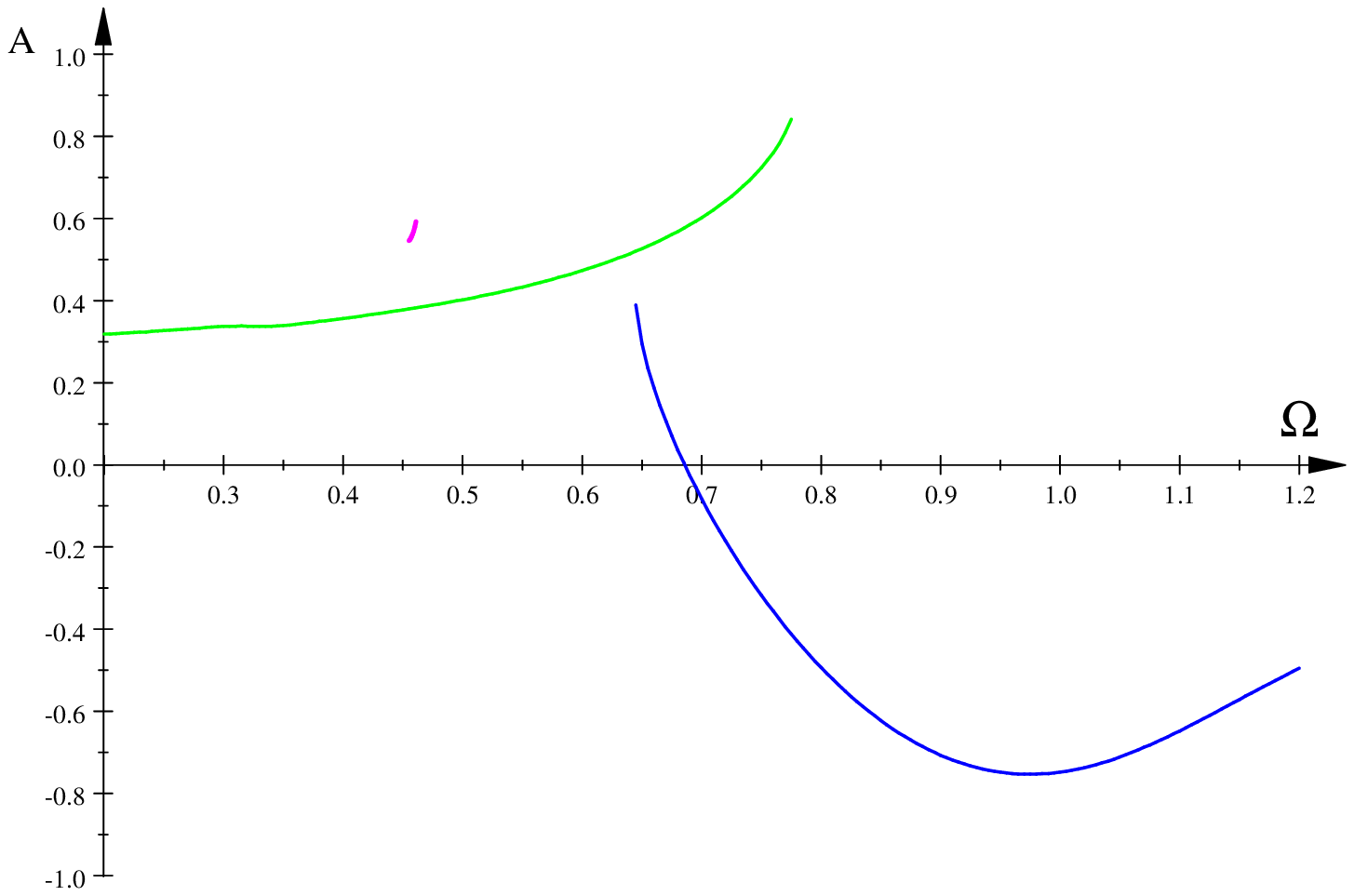}
\caption{Bifurcation diagrams, $f=0.2990$ left -- before the isolated point is formed, $f=0.2991$ right -- just 
after formation of a new branch of solution (magenta). Colors (green, blue, magenta) correspond to colors in Fig. (\ref{F3}).}
\label{F4}
\end{figure}
\begin{figure}[t!]
\center
\includegraphics[width=9cm, height=6cm]{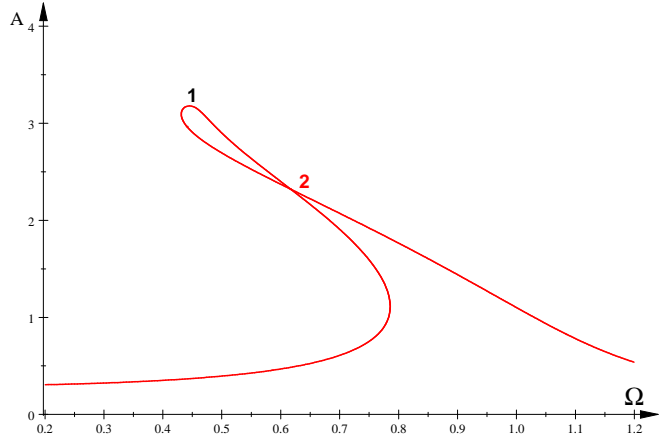}
\caption{Amplitude profile, $h=0.2,$ $c_{3}=-0.2,$ $c_{5}=1.15\times 10^{-2},$ $f=290\,089$}
\label{F5}
\end{figure}

There are three characteristic points in Fig. (\ref{F3} -- solutions of Eq. (\ref{f(Z)}). 
There is a cusp of the degenerate amplitude profile (red), singular with multiplicity $3$ -- a triple solution of Eq. (\ref{f(Z)}). 
Moreover there are two interesting points on the amplitude with isolated point (green, blue, magenta): 
an  isolated point -- singular with multiplicity $2$ (magenta dot) and a local maximum with multiplicity $1$ -- non-singular 
and the corresponding bifurcation diagrams in Fig. \ref{F4}  document indeed bifurcation due to 
metamorphosis of the amplitude profiles.
The amplitude profile for the self-intersection is shown in Fig. \ref{F5} 
\newpage

\begin{figure}[h!]
\center
\includegraphics[width=6cm, height=6cm]{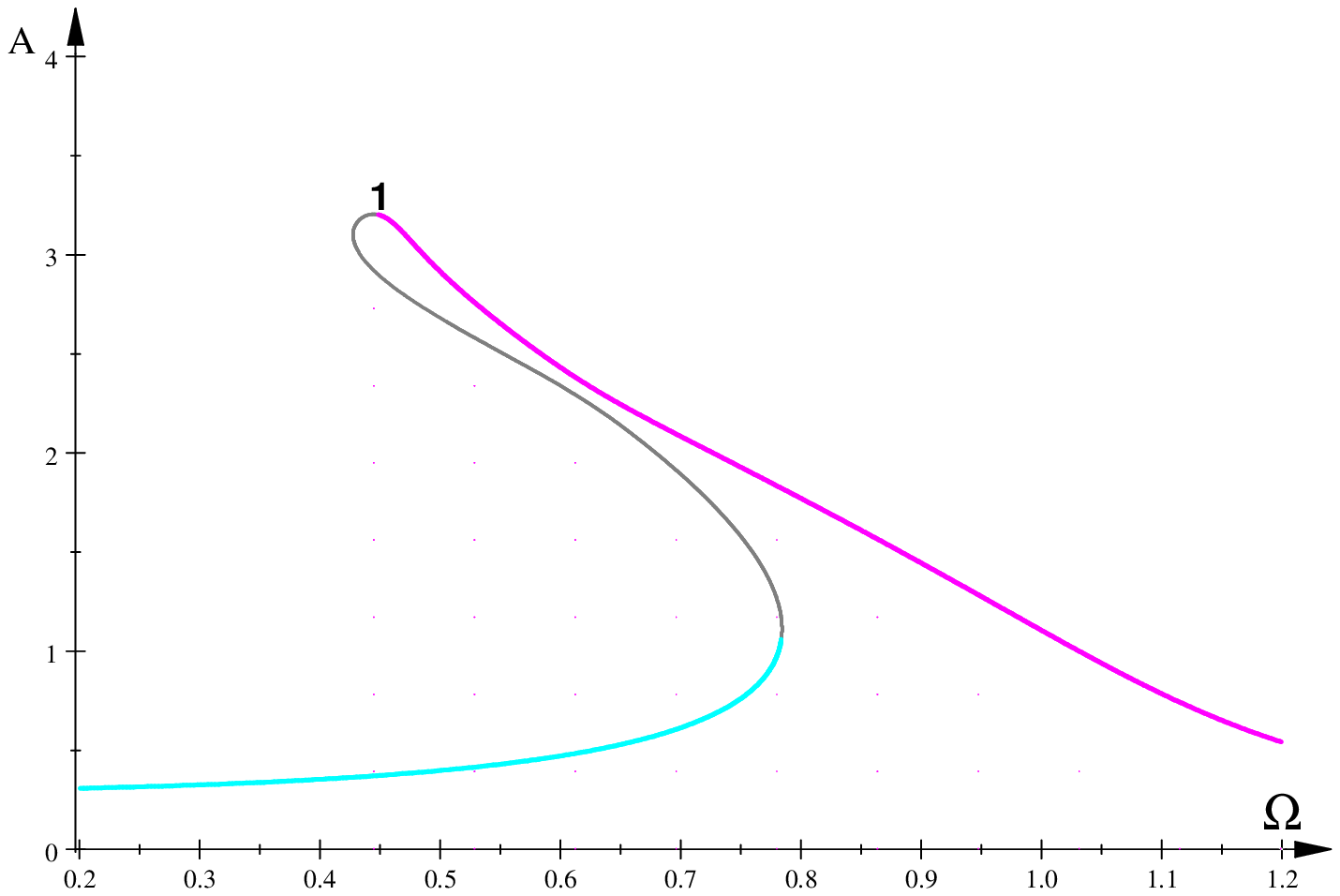}
\includegraphics[width=6cm, height=6cm]{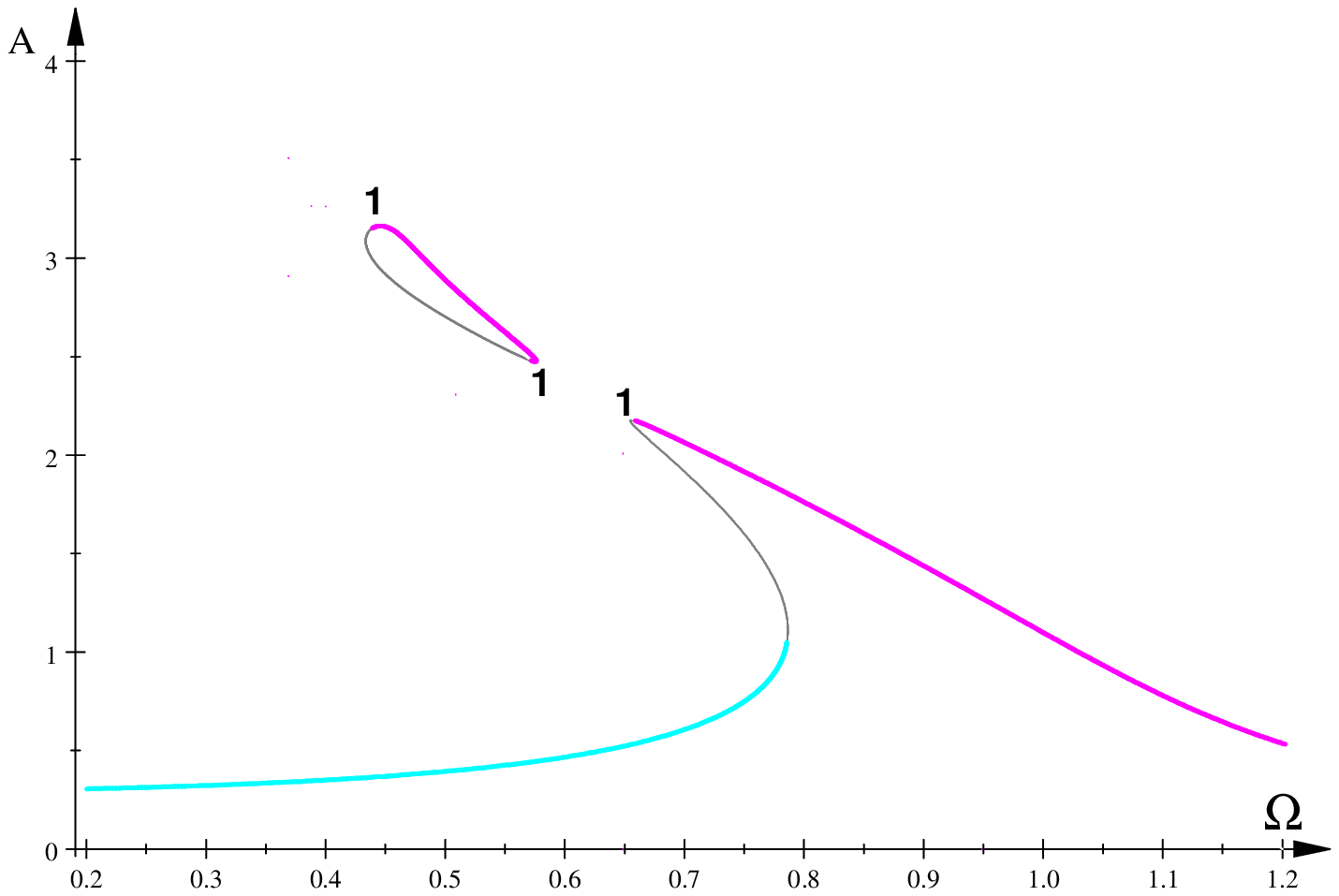}
\caption{Amplitude profiles, $f=0.292$ (left, bofore the self-intersection), $f=0.289$ (right, after the self intersection).}
\label{F6}
\end{figure}

\noindent and two amplitude profiles near the intersection are also shown in Fig. \ref{F6}.

The corresponding bifurcation diagrams document indeed bifurcation due to 
metamorphosis of the amplitude profiles.
\begin{figure}[h!]
\center
\includegraphics[width=6cm, height=6cm]{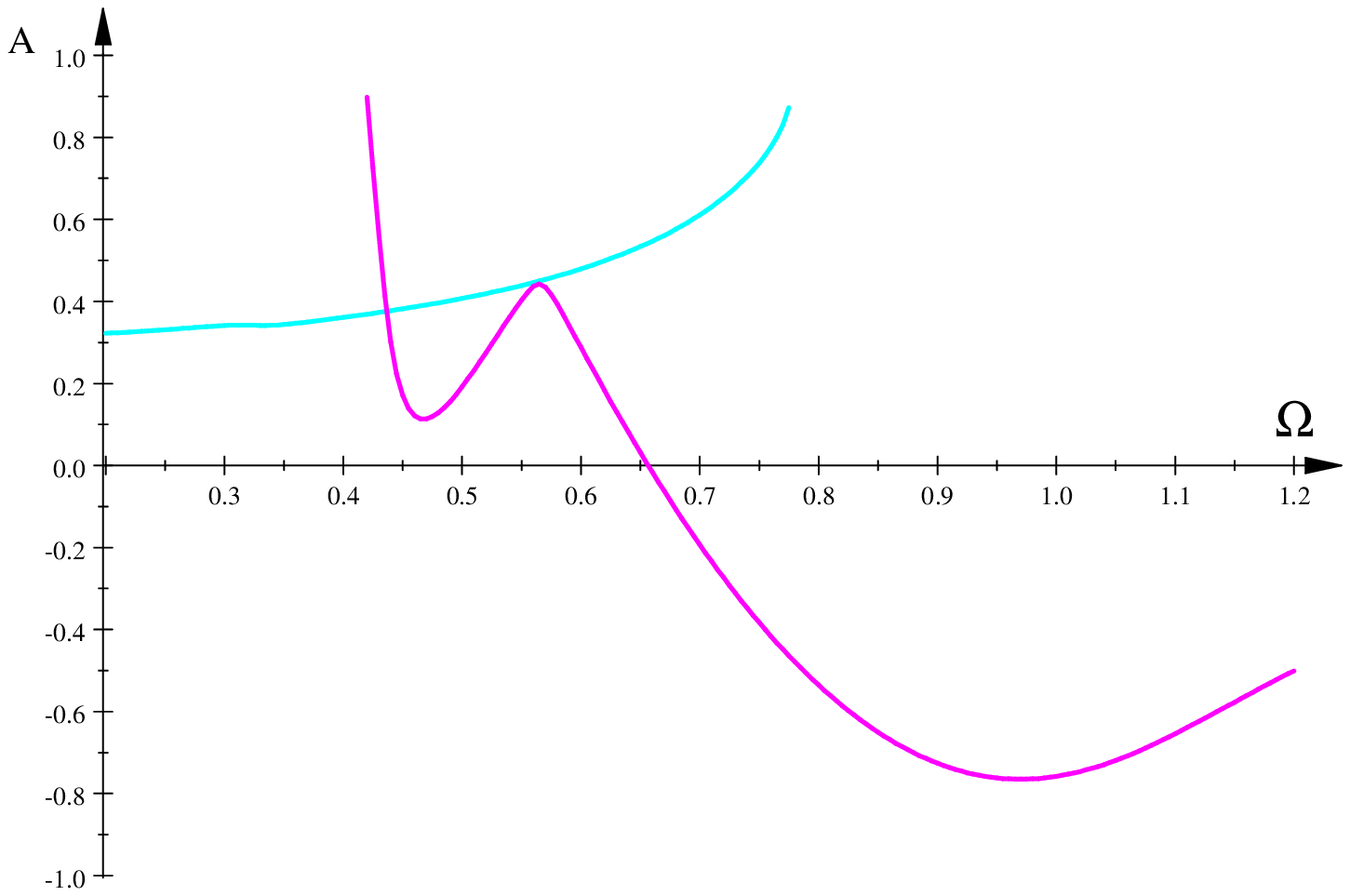}
\includegraphics[width=6cm, height=6cm]{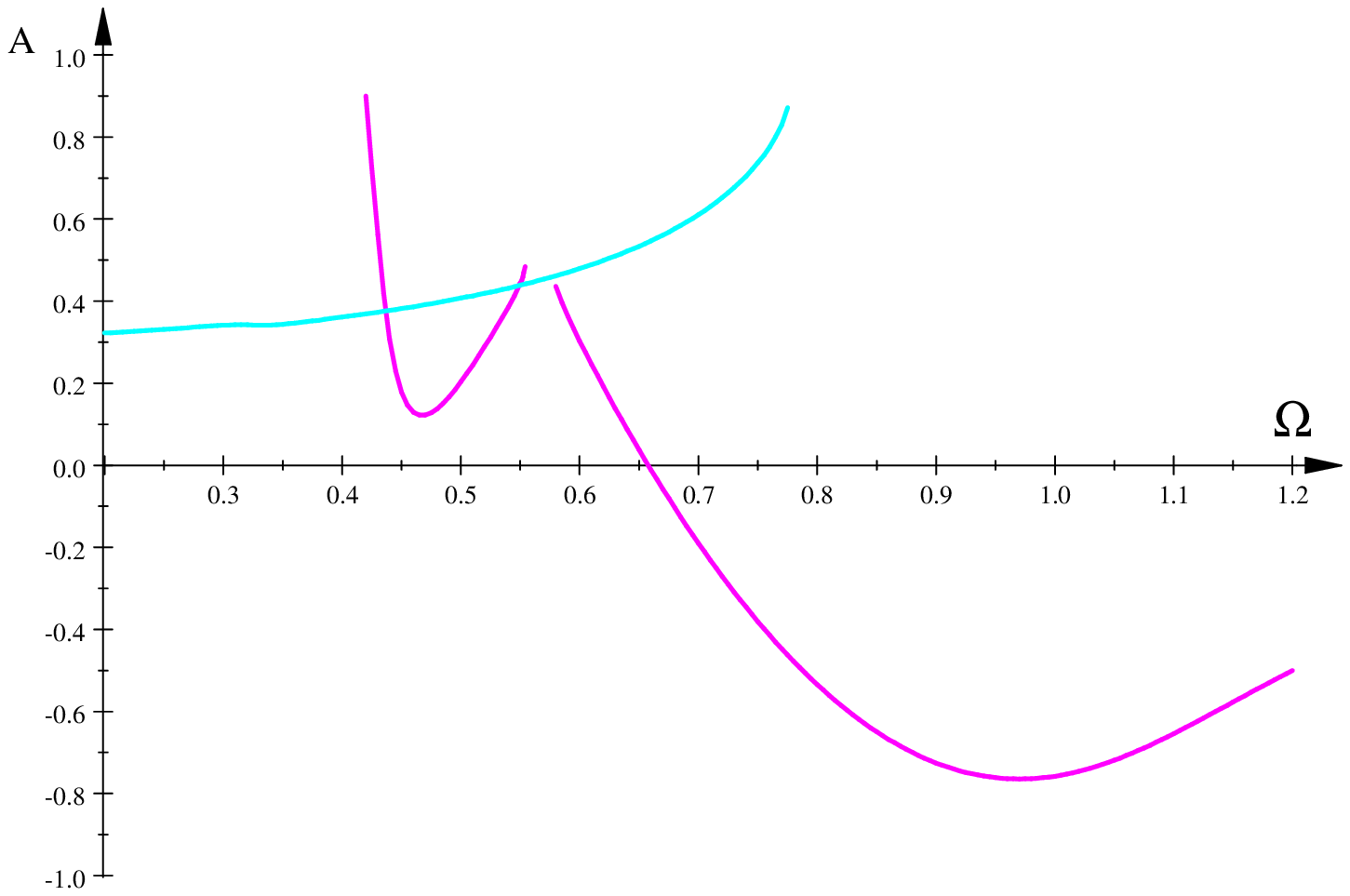}
\caption{Bifurcation diagrams, $f=0.3026$ left, $f = 0.3025$ right.}
\label{F7}
\end{figure}

\newpage
The bifurcation, discontinuity of the magenta line in Fig. \ref{F7}, appears in numerical integration 
of the Duffing equation for $0.3025<f<0.3026$ in good agreement with the predicted value 
($f=0.290\,089$).

\section{Metamorphoses and bifurcations}
\label{Classification}
\subsection{Non-singular metamorphoses}
\begin{figure}[h!]
\center
\includegraphics[width=9cm, height=6cm]{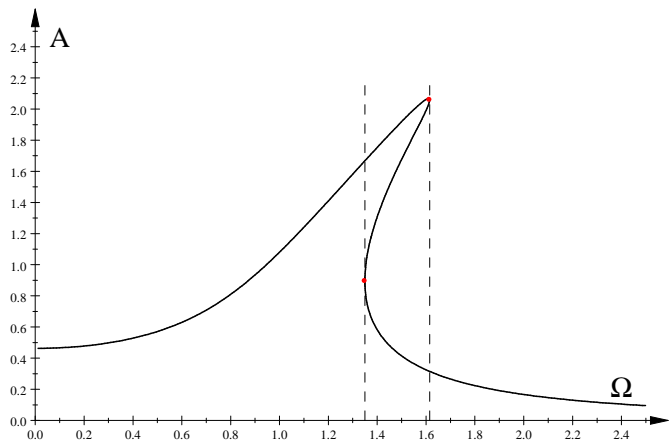}
\caption{Jump phenomenon for the Duffing equation, $h=0.15$, $c_{3}=0.5$, $f=0.5$.}
\label{FD}
\end{figure}

It should be stressed that metamorphoses, i.e. changes of differential properties of asymptotic solutions, 
equivalent to bifurcations, occur also at nonsingular, yet critical, points of the amplitude-response equation. To show this 
let us consider the jump phenomenon for the Duffing equation \cite{Kalmar2011,Holmes1976}, see Fig. \ref{FD}. 
In two points marked with red dots, $\left( 1.\,349,\ 0.894\right) $, $\left( 1.\,615,\ 2.\,058\right) $, 
which are solutions of Eqs. (\ref{CONDITIONS}), 
 metamorphoses occur -- 
a number of branches of the asymptotic solution is changed \cite{Holmes1976}.  Real solutions appear for 
 $h<0.304$, $c_{3}=0.5$, $f=0.5$, see \cite{Holmes1976} for analytical condition.
It was determined by  Kalm\'{a}r-Nagy and Balachandran that this metamorphoses follows from a differential 
condition \cite{Kalmar2011}:
\begin{equation}
\frac{d\Omega }{dA}=0  \label{K-N,B}
\end{equation}
where the amplitude-response curve for the cubic Duffing equation is $F\left( \Omega ,A\right) =L_{3}\left( \Omega ^{2},A^{2}\right)
=h^{2}A^{2}\Omega ^{2}+A^{2}\left( \Omega ^{2}-1-\frac{3}{4}
c_{3}A^{2}\right) ^{2}-f^{2} = 0$
and is equivalent to the saddle-node bifurcation 
since one eigenvalue of the Jacobian matrix is zero, while another is negative. These metamorphoses can be also 
referred to as vertical tangencies of the response curve \cite{Kalmar2011,Nayfeh2008}. 

The condition (\ref{K-N,B}) can be formulated within the framework of
implicit function theorem.
Consider implicit amplitude response curve:%
\begin{equation}
F\left( \Omega ,\ A\right) =0.  \label{L}
\end{equation}%
Let $\Omega =f\left( A\right) $. Then:%
\begin{equation}
\frac{\partial F}{\partial \Omega }\dfrac{d\Omega }{dA}+\frac{\partial F}{%
\partial A}\dfrac{dA}{dA}=0,  \label{ChainRule}
\end{equation}%
and%
\begin{equation}
\dfrac{d\Omega }{dA}=f^{\prime }\left( A\right) =-\frac{\frac{\partial F}{%
\partial A}}{\frac{\partial F}{\partial \Omega }},\quad \left( \tfrac{%
\partial F}{\partial \Omega }\neq 0\right) ,  \label{Derivative}
\end{equation}%
see Section 11.5 in \cite{Stewart2009}.

Therefore, critical points of the function $\Omega =f\left( A\right) $, i.e.
vertical tangencies, which follow from the Kalm\'{a}r-Nagy and Balachandran
condition $\dfrac{d\Omega }{dA}=f^{\prime }\left( A\right) =0$, fulfill an
equivalent set of equations: 
\begin{subequations}
\label{CONDITIONS}
\begin{eqnarray}
F\left( \Omega ,\ A\right) &=&0,  \label{c1} \\
\frac{\partial F\left( \Omega ,\ A\right) }{\partial A} &=&0.  \label{c2}
\end{eqnarray}
\end{subequations}

\subsection{Singular metamorphoses}
Singular points of the amplitude-response curves fulfill equations (\ref{SINGULAR}). 

\noindent The first and third of these 
equations are conditions for vertical tangencies (\ref{CONDITIONS}), associated with saddle-node 
bifurcations \cite{Kalmar2011}. Therefore, because of the additional condition (\ref{S2}), 
singular points of the amplitude-response curves lead to more complicated metamorphoses of these curves, 
discussed in Sections \ref{cubic}, \ref{cubic-quintic}.

On the other hand, also singular metamorphoses, due to Eq. (\ref{S3}), are associated with saddle-node 
bifurcations. 
A connection between metamorphoses and bifurcations is revealed by computation of 
determinant of the Jacobian matrix $\mathbb{J}$. It follows from Eqs. (\ref{J}), (\ref{SECULAR2}) 
that
\begin{equation}
\det \left( \mathbb{J}\right) =\frac{1}{4X}\frac{\partial L}{\partial Y},
\label{determinant}
\end{equation}
where, for simplicity, the determinant is written in variables $X=\Omega ^{2}$, $Y=A^{2}$. 

It now follows that condition $\frac{\partial L\left( X,\ Y\right) }{\partial Y}=0$ $\left( \text{or } 
\frac{\partial L\left( \Omega ^{2},\ A^{2}\right) }{\partial A}=0\right) $, defining a vertical tangency, 
 is equivalent to 
vanishing of the determinant $\det \left( \mathbb{J}\right) $, and this means that at least one eigenvalue 
of the Jacobian matrix is zero, indicating  a bifurcation. 

We have computed another eigenvalue of the Jacobian matrix and in all cases considered it was equal $-h$. 
Therefore all bifurcations described in Sections \ref{cubic}, \ref{cubic-quintic} 
 are saddle-node bifurcations of co-dimension one.

\section{Summary and discussion}
\label{Summary}
In this work, we have studied changes of differential properties -- metamorphoses --  of amplitude response curves 
for the generalized Duffing equation with polynomial nonlinearities  (\ref{gen-Duff-1}). 
We have demonstrated that metamorphoses are due to formation of singular points on amplitude profiles 
(the case of singular metamorphoses) and due to formation of critical points on these curves (the case of 
non-singular metamorphoses). The non-singular case, first described in \cite{Kalmar2011} 
as due to formation of vertical tangent points,  leads to important jump phenomena \cite{Kalmar2011,Holmes1976}. 
We discuss singular and non-singular metamorphoses and associated bifurcations in Section \ref{Classification}.

More precisely, we have derived formulae for singular points and bifurcation sets 
of the amplitude response equation  for the generalized Duffing equation (\ref{gen-Duff-1}), 
$n=3,5,7,\ldots$.
We have described singular metamorphoses for the cubic Duffing equation ($n=3$) and 
the cubic-quintic Duffing equation ($n=5$). 

It is interesting that there is a singular point in the case of the standard Duffing equation, see 
the corresponding metamorphoses of the amplitude profile and change of dynamics,
 cf. Figs. \ref{F1}, \ref{F2}. 
However, the set of parameters for which such points exist is rather small and can be easily 
overlooked.

In the cubic-quintic equation, $n=5$, there are degenerate points for $c_3<0$ and two 
infinite sets of self-intersections and isolated points in the neighbourhoods of these points, see Figs. 
\ref{F3}, \ref{F5}, \ref{F6}. Bifurcations diagrams show indeed changes of dynamics -- birth of 
new branches of solutions, Fig. \ref{F4}, and rupture of existing branches, Fig. \ref{F7}. 
\medskip

Summing up, knowledge of metamorphoses, non-singular, as well as singular, permits prediction of changes of 
dynamics  such as jump phenomena (due to vertical tangent points), birth of new branches of solutions (due to 
isolated points), and rupture of existing branches of solutions leading to gaps in bifurcation diagrams (due to 
self-intersections). Knowledge of degenerate singular points pinpoints regions in the parameter space with 
families of isolated points and self-intersections where very complicated dynamical phenomena can occur. 

There is an alternative approach to singular points of amplitude profiles when the implicit function 
can be disentangled, see \ref{A} and \ref{B}. 

In the  \ref{A} we show that the implicit equation $F\left( A,\Omega \right) =0$ can be solved for the
cubic-quintic Dufing equation resulting in explicit expression for two
branches in form $\Omega _{\pm }=f_{\pm }\left( A\right)$ (actually equation (\ref{Amplitude}) can be 
solved for $\Omega$ for any $n$). 
Condition of intersection of these branches, $f_{+}\left( A\right)
=f_{-}\left( A\right)$ leads to dynamically interested points, non-singular as well as singular. 
It is important, that this condition, Eq. (\ref{condition}),  is equivalent to the equation  (\ref{f(Z)}), 
derived within a more general approach.
Accordingly, all these points are single or multiple solutions of Eq.  (\ref{f(Z)}). 
Within this approach we describe metamorphoses of the amplitude profiles in a more detailed way.

We compare  in the \ref{B}  the computed amplitude profiles for the cubic-quintic Duffing equation 
with analogous solutions obtained by Karahan and Pakdemirli \cite{Karahan2017} 
within the Multiple Scales Lindstedt Poincar\'{e} (MSLP) approach, obtaining a qualitative 
agreement.
 
\appendix{}
\section{Cubic-quintic Duffing equation: alternative approach to singular points of amplitude profiles}
\label{A}
Solving Eq. (\ref{L(3-5)}) for $\Omega $ we get two positive solutions: 
\begin{subequations}
\label{SOL}
\begin{eqnarray}
\Omega _{\pm } &=&\tfrac{1}{4}\sqrt{\tfrac{2}{A}}\sqrt{f\left( A\right) \pm 2%
\sqrt{g\left( A\right) }}  \label{branches} \\
f\left( A\right)  &=&5c_{5}A^{5}+6c_{3}A^{3}+4A\left( 2-h^{2}\right) 
\label{function1} \\
g\left( A\right)  &=&-10c_{5}A^{6}h^{2}-12c_{3}A^{4}h^{2}+4A^{2}h^{2}\left(
h^{2}-4\right) +16f^{2}  \label{function-2}
\end{eqnarray}

The branches (\ref{branches}) intersect for 
\end{subequations}
\begin{equation}
g\left( A\right) =-10c_{5}A^{6}h^{2}-12c_{3}A^{4}h^{2}+4A^{2}h^{2}\left(
h^{2}-4\right) +16f^{2}=0  \label{condition}
\end{equation}
It follows that $g\left( A\right) =-2f_{5}\left( A^{2}\right) $, cf. Eq. (\ref{f(Z)}), 
and thus conditions $g\left( A\right) =0$ and $f_{5}\left( A^{2}\right) =0$
are equivalent.

\begin{figure}[h!]
\center
\includegraphics[width=6cm, height=6cm]{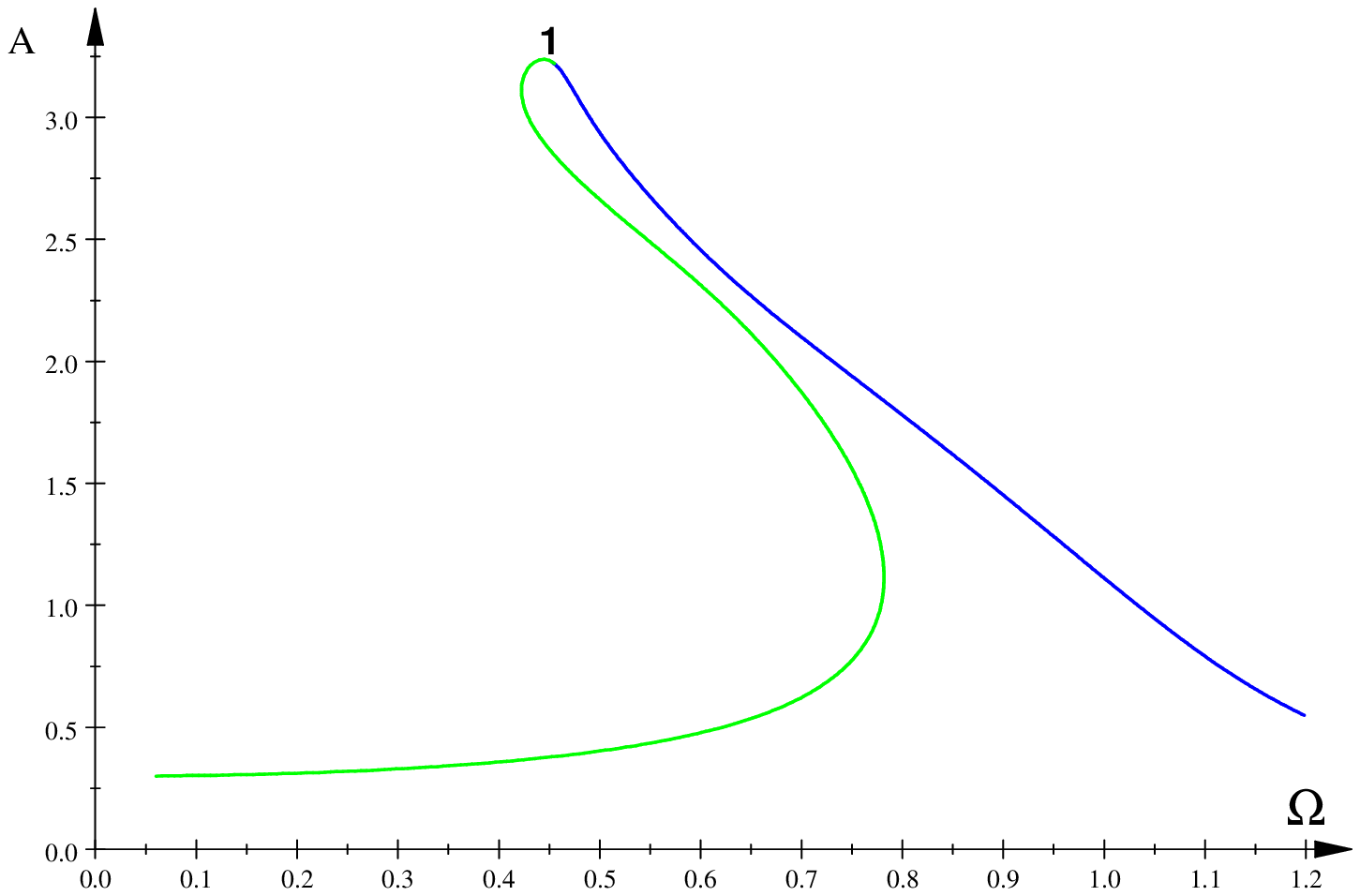}
\includegraphics[width=6cm, height=6cm]{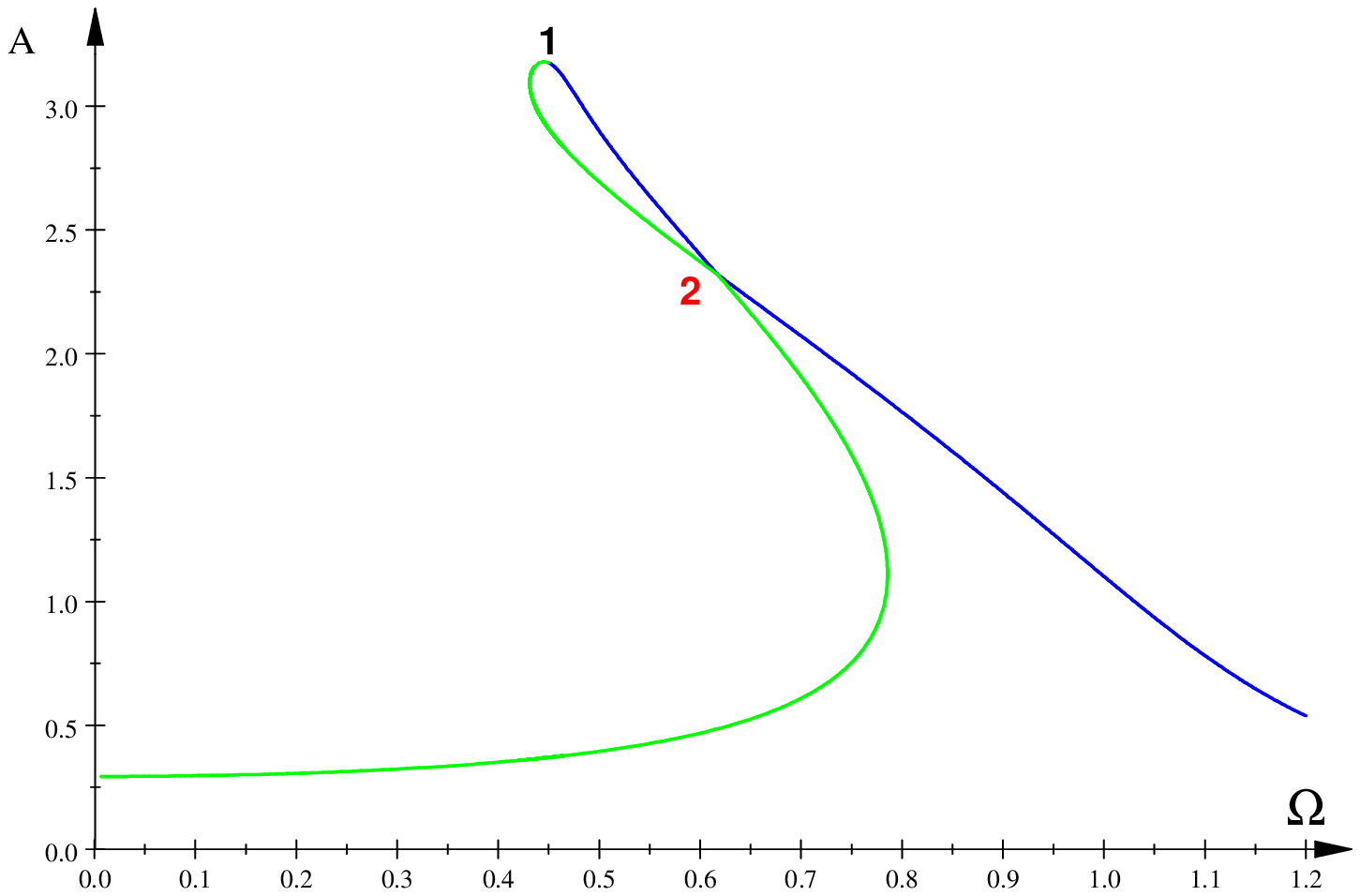}
\caption{Amplitude profiles: $f=0.295$ left, $f=0.290\,089$ right}
\label{F8}
\end{figure}
\begin{figure}[t!]
\center
\includegraphics[width=6cm, height=6cm]{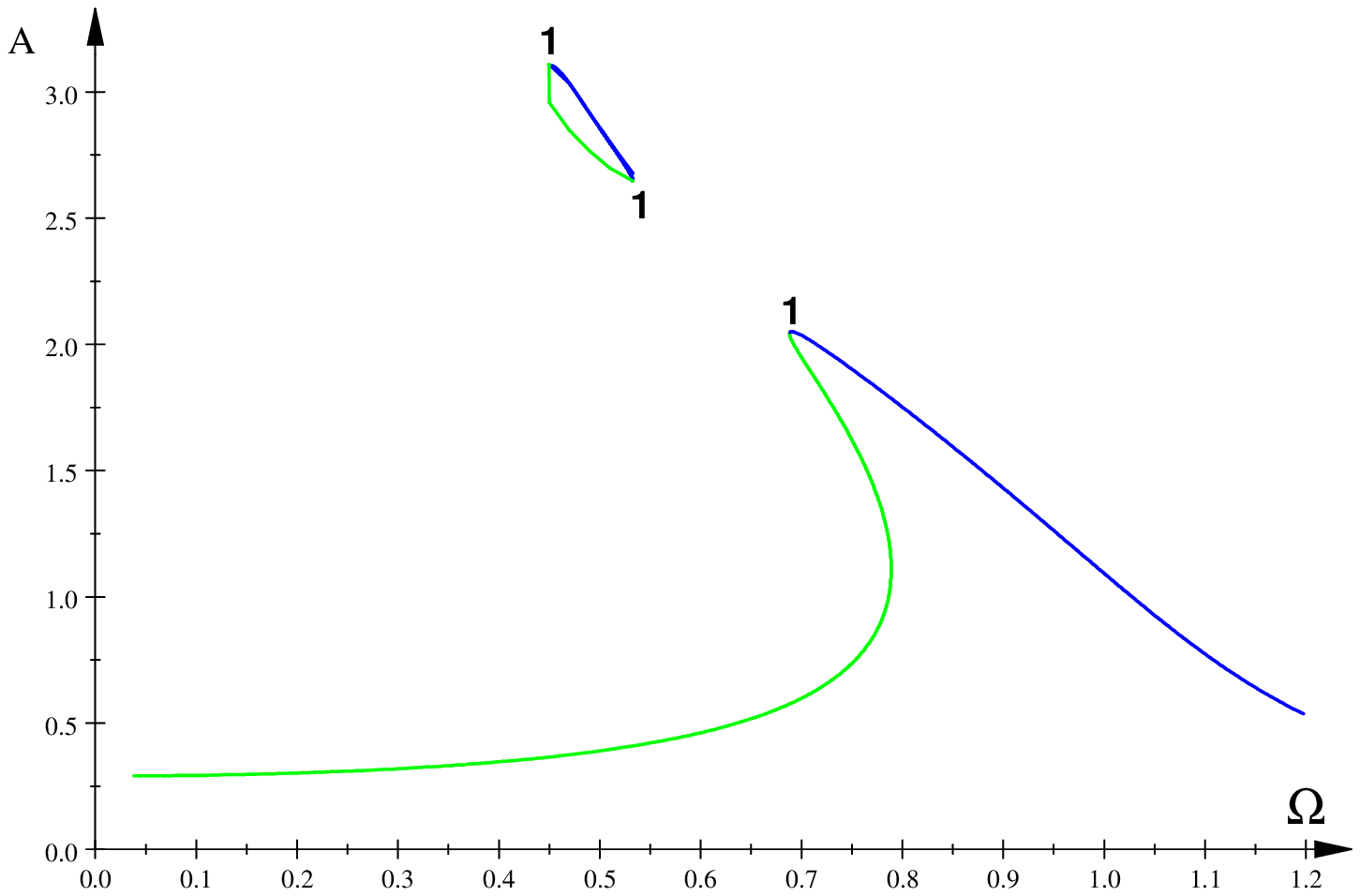}
\includegraphics[width=6cm, height=6cm]{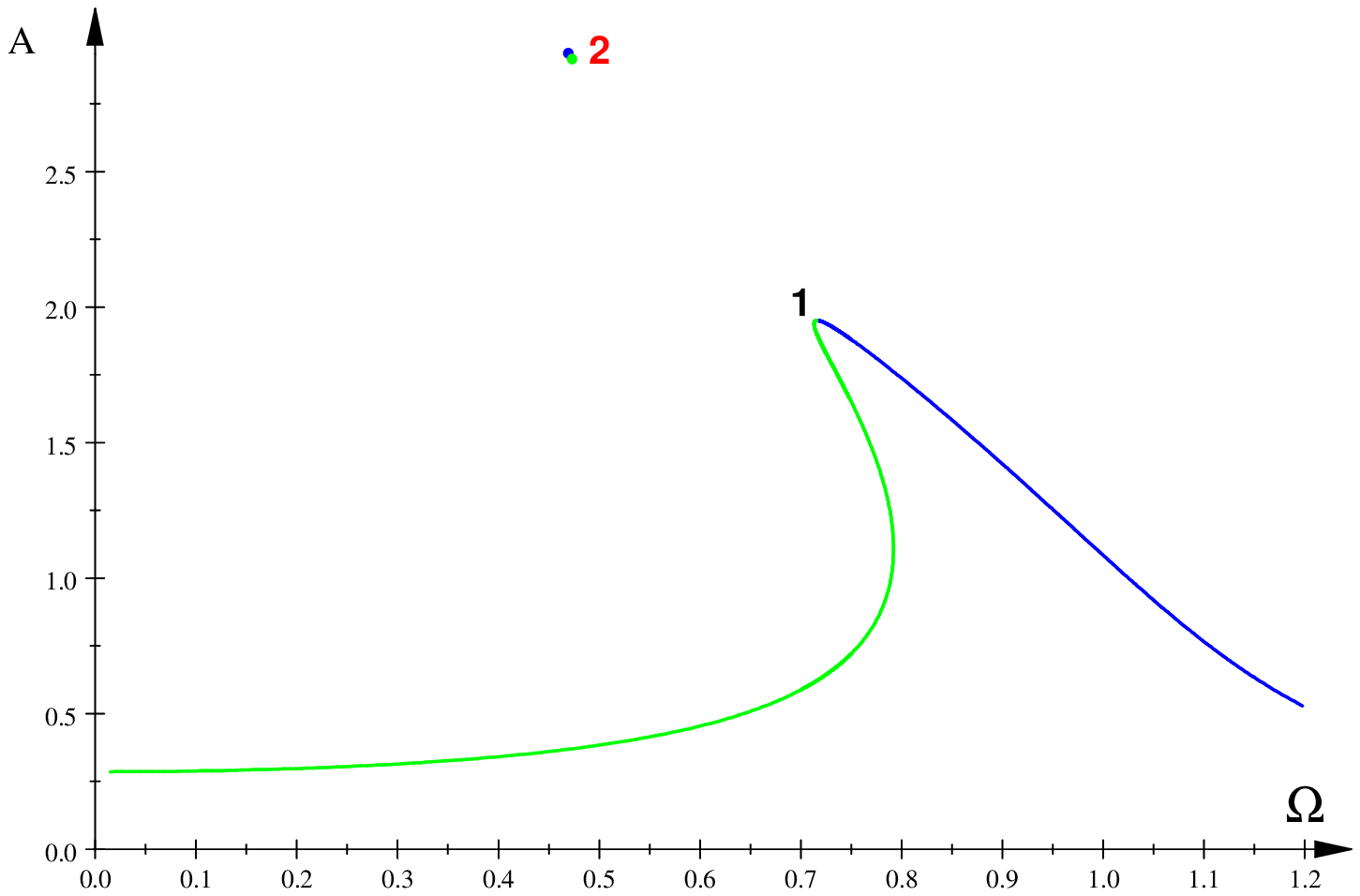}
\caption{Amplitude profiles: $f=0.286$ left, $f=0.282\,25$ right}
\label{F9}
\end{figure}

In Figs. \ref{F8}, \ref{F9} transition from non-singular amplitude profile to self-intersection, 
to nonsingular profile, and to an isolated point is shown for  $h=0.2$, $c_{3}=-0.2$, $ c_{5}=0.0115$, 
and values of $f$ shown in the Figures.

Branches $\Omega _{+}$ and $\Omega _{-}$ are colored blue and green,
respectively, black and red digits in the Figures denote multiplicity of the intersections of the branches 
(i.e. multiplicity of solutions of Eq. (\ref{condition})) and correspond to non-singular and singular cases, 
respectively.

\newpage

Note that this approach works for the generalized Duffing equation (\ref{gen-Duff-2}) for any $n$ 
since the amplitude equation (\ref{Amplitude}) can be solved for $\Omega$ for an arbitrary $n$. 

\section{Comparison with asymptotic solution from Ref. \cite{Karahan2017}}
\label{B}
The cubic-quintic Duffing equation was also solved by application of
Multiple Scales Lindstedt Poincar\'{e} (MSLP) method by Karahan and Pakdemirli, see 
Eqs. (89), (90) in \cite{Karahan2017}. The authors computed explicit solution 
as a function $\Omega =f\left( A\right)$ consisting of two branches:
\begin{subequations}
\label{KARAHAN}
\begin{eqnarray}
\frac{\Omega _{\pm }}{\omega } &=&1+\varepsilon ^{2}\left( \dfrac{3\alpha
^{3}}{256\omega ^{4}}A^{4}+\dfrac{10\beta }{32\omega ^{2}}A^{4}\pm \frac{1}{2%
}\sqrt{\dfrac{F^{2}}{A^{2}\omega ^{4}}-\left( \dfrac{\mu }{\omega }\right)
^{2}}\right)   \label{Ksol1} \\
\omega  &=&\sqrt{1+\varepsilon \frac{3}{4}\alpha A^{2}}  \label{Ksol2}
\end{eqnarray}
\end{subequations}
Parameters used in \cite{Karahan2017} and our parameters are related:
\begin{equation}
h=\varepsilon ^{2}\mu ,\ c_{3}=\varepsilon \alpha ,\ c_{5}=\varepsilon
^{2}\beta ,\ f=\varepsilon ^{2}F  \label{parameters}
\end{equation}
Condition for intersection of these branches is:
\begin{equation}
\dfrac{F^{2}}{A^{2}\omega ^{4}}-\left( \dfrac{\mu }{\omega }\right) ^{2}=0.
\label{intersection}
\end{equation}

In Figures below $\varepsilon =1$, $\mu =h=0.2$, $\alpha =c_{3}=-0.2$,
 $\beta =c_{5}=0.0115$.
Red digits indicate multiplicity of intersections of the branches (multiplicity of solutions of Eq. (\ref{intersection}), 
branches $\omega _{+}$ and $\omega _{-}$ are colored blue and green, respectively. 

Intersection of multiplicity $2$ appears for $F=f=0.258\,199$ what can be compared with the 
value obtained by numerical integration of the cubic-quintic Duffing equation $0.3025<f<0.3026$ 
(we have obtained from the KBM implicit function (\ref{L(3-5)}) value $f=290\,089$). 

\begin{figure}[h!]
\center
\includegraphics[width=6cm, height=6cm]{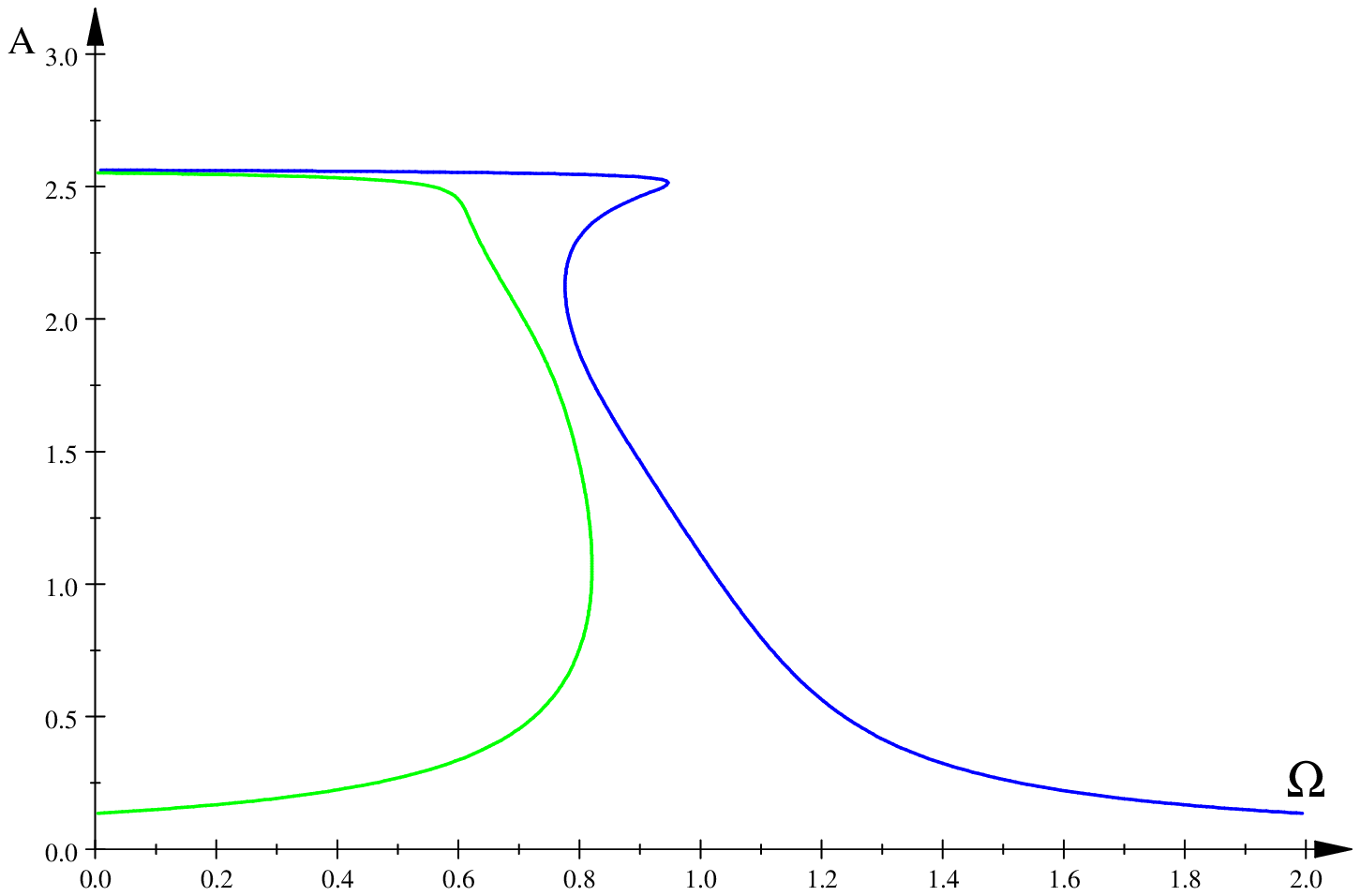}
\includegraphics[width=6cm, height=6cm]{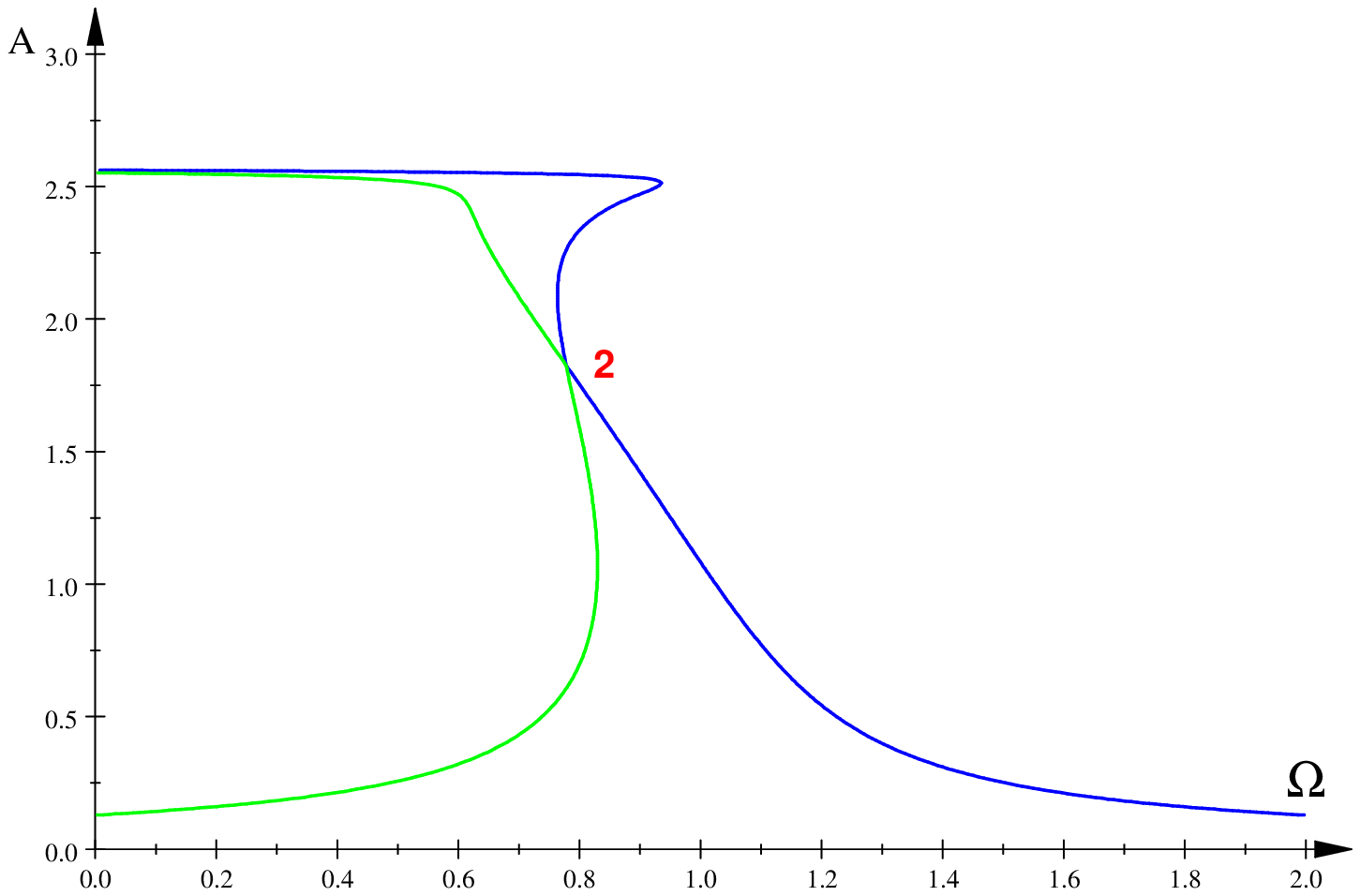}
\caption{Amplitude profiles: $F=f=0.27$ left, $F=f=0.258\,199$ right}
\label{K1}
\end{figure}

\begin{figure}[h!]
\center
\includegraphics[width=6cm, height=6cm]{Karahan-2-new.eps}
\includegraphics[width=6cm, height=6cm]{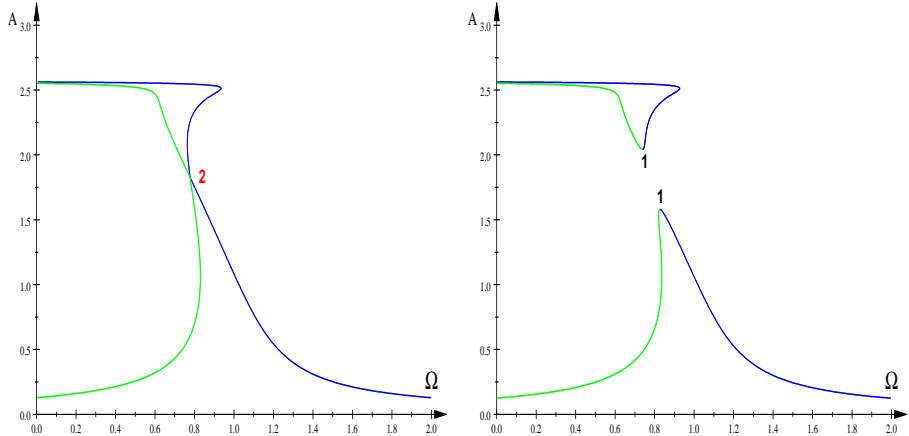}
\caption{Amplitude profiles:  $F=f=0.258\,199$ left, $F=f=0.25$ right}
\label{K2}
\end{figure}

Alternatively, we can compute singular points writing Eq. (\ref{Ksol1}) in form 
$f\left( \Omega ,A\right) =\pm \sqrt{g\left( \Omega ,A\right) }$ to obtain 
 implicit equation of form $K\left(
\Omega ,A\right) =f^{2}\left( \Omega ,A\right) -g\left( \Omega ,A\right) =0$
and applying standard equations:

\begin{subequations}
\label{KSING1}
\begin{align}
K\left( \Omega ,A\right) & =0,  \label{KSing1a} \\
\frac{\partial K\left( \Omega ,A\right) }{\partial \Omega }& =0,
\label{KSing1b} \\
\frac{\partial K\left( \Omega ,A\right) }{\partial A}& =0.  \label{KSing1c}
\end{align}
\end{subequations}
We can demonstrate within this approach that functions (\ref{KARAHAN}) do not have 
neither degenerate nor isolated points as singular solutions. It seems, however, that such points will be 
present if the MSLP solution contains higher-order terms. 

\section{Computational details}
\label{Details}
 Nonlinear polynomial equations were solved numerically using
the computational engine Maple 4.0 from the Scientific WorkPlace 4.0. 
Figures were plotted with the
computational engine MuPAD 4.0 from Scientific WorkPlace 5.5. Curves shown in
bifurcation diagrams in Figs. \ref{F2}, \ref{F4}, \ref{F7} were computed by
integrating numerically Eq. (\ref{gen-Duff-2}), $n=3, 5$, running DYNAMICS, program written by
Helena E. Nusse and James A. Yorke \cite{Nusse1997}, and our own programs
written in Pascal.

\end{document}